\documentclass[fleqn,11pt,twoside]{article}
\usepackage{amsthm}
\usepackage{amsmath}
\usepackage{graphicx}
\usepackage{lscape}
\usepackage{latexsym}
\usepackage[all]{xy}
\makeatletter

\newcommand{\Name}[1]{\begin{flushleft}
                       \LARGE \bf #1
                       \end{flushleft}\vspace{-3mm}}

\newcommand{\Author}[1]{\begin{flushleft}
                       \it #1 \end{flushleft}}

\newcommand{\Address}[1]{\begin{flushleft}
                       \it #1 \end{flushleft}}

\newcommand{\FirstPageHead}[5]{
\begin{flushleft}
\raisebox{8mm}[0pt][0pt]
{\footnotesize \sf
\parbox{150mm}{ \qquad
 #1 #2 #3 
#4\hfill {\sc #5}}}\vspace{-13mm}
\end{flushleft}}

\newcommand{\evenhead}{Author \ name}
\newcommand{\oddhead}{Article \ name}

\renewcommand{\@evenhead}{
\hspace*{-3pt}\raisebox{-15pt}[\headheight][0pt]{\vbox{\hbox to \textwidth
{\thepage \hfil \evenhead}\vskip4pt \hrule}}}
\renewcommand{\@oddhead}{
\hspace*{-3pt}\raisebox{-15pt}[\headheight][0pt]{\vbox{\hbox to \textwidth
{\oddhead \hfil \thepage}\vskip4pt\hrule}}}
\renewcommand{\@evenfoot}{}
\renewcommand{\@oddfoot}{}

\long\def\@makecaption#1#2{%
  \vskip\abovecaptionskip
  \sbox\@tempboxa{\small \textbf{#1.}\ \ #2}%
  \ifdim \wd\@tempboxa >\hsize
    {\small \textbf{#1.}\ \ #2}\par
  \else
    \global \@minipagefalse
    \hb@xt@\hsize{\hfil\box\@tempboxa\hfil}%
  \fi
  \vskip\belowcaptionskip}

\newcommand{\JNMPnumberwithin}[3][\arabic]{%
  \@ifundefined{c@#2}{\@nocounterr{#2}}{%
    \@ifundefined{c@#3}{\@nocnterr{#3}}{%
      \@addtoreset{#2}{#3}%
      \@xp\xdef\csname the#2\endcsname{%
        \@xp\@nx\csname the#3\endcsname .\@nx#1{#2}}}}%
}

\newcommand{\resetfootnoterule} {
  \renewcommand\footnoterule{%
  \kern-3\p@
  \hrule\@width.4\columnwidth
  \kern2.6\p@}
}

\renewcommand{\footnoterule}{}

\newcommand{\be}{\begin{equation}}
\newcommand{\ee}{\end{equation}}
\newcommand{\ba}{\hspace*{-5pt}\begin{array}}
\newcommand{\ea}{\end{array}}
\newcommand{\p}{\partial}

\makeatother

\numberwithin{equation}{section}
\theoremstyle{definition}

\renewcommand{\ba}{\begin{array}}
\renewcommand{\ea}{\end{array}}
\newcommand{\beg}{\begin{eqnarray}}
\newcommand{\eeq}{\end{eqnarray}}
\newcommand{\bg}{\begin{eqnarray*}}

\newcommand{\ed}{\end{eqnarray*}}
\newcommand{\n}{\newline\hfill}
\newcommand{\nn}{\nonumber}

\renewcommand{\p}{\partial} 
 
\newcommand{\notlhd}{\lhd\kern-.8em{/}\ } 
\newcommand{\notexist}{\ \exists\kern-.5em{\raise.1em\hbox{/}}\ }

\newcommand{\pde}[2]{\frac{\p #1}{\p #2}}

\newcommand{\inp}{{\mbox{\vbox{\hrule width0ex\hbox{\vrule
 height0ex\kern3.8pt
\vbox{\kern2.5pt}\kern3.8pt \vrule height1.6ex}
\hrule width1.6ex}}}}
\newcommand{\bm}[1]{\mbox{\boldmath$#1$}}

\begin{document}

\renewcommand{\evenhead}{N Euler and M Euler}
\renewcommand{\oddhead}{The converse problem of multipotentialisation}


\thispagestyle{empty}

\begin{flushleft}
\footnotesize \sf
\end{flushleft}

\FirstPageHead{\ }{\ }{\ }
{ }{{
{ }}}

\Name{The converse problem for the multipotentialisation of 
evolution equations and systems}

\Author{\bf Norbert Euler and Marianna Euler}



\Address{
Department of Mathematics,  
Lule\aa\ University of Technology \\
SE-971 87 Lule\aa, Sweden\\
Emails: norbert@ltu.se; marianna@ltu.se;
}

\vspace{1cm}

\noindent
{\bf Abstract}: 
We propose a method to identify and classify evolution equations and
systems that can be multipotentialised in given target equations or
target systems.
We refer to this as the {\it converse problem}. Although we mainly 
study a method for $(1+1)$-dimensional equations/system, we do also
propose an extension of the methodology to higher-dimensional 
evolution equations. An important point is that the proposed 
converse method allows one to identify certain types of
auto-B\"acklund transformations for the equations/systems. In this respect we
define the {\it triangular-auto-B\"acklund transformation}
and derive its connections to the converse problem. Several explicit 
examples are given. In particular we investigate a class of linearisable
third-order evolution equations, a fifth-order
symmetry-integrable evolution equation as well as 
linearisable systems.

\tableofcontents

\vspace{1cm}


\section{Introduction}

Potentialisations of evolution equations provides a natural way to
study special types of nonlocal symmetries for partial differential
equations and systems, known as potential 
symmetries \cite{Bluman+Kumei}. 
In some cases it is possible to apply
the potentialisation process again on the derived potential equations
themselves, which is known as the mulipotentialisation process.
This procedure of multipotentialisation was applied
in  \cite{Euler_nonlocal_2009} and \cite{Euler_NK_2009}
to investigate higher-degree potential symmetries, nonlocal
transformations, nonlocal conservation laws, as well as 
iterating-solution formulae; all of which 
were derived as a direct consequence of a systematic
multipotentialisation 
of the equations.
In \cite{Euler_nonlocal_2009} 
we introduced higher-degree potential symmetries for the Burgers'-
\cite{EEP} and 
Calogero-Degasperis-Ibragimov-Shabat hierarchies \cite{PEE}
and derived the
nonlocal linearisation transformations by means of a 
multipotentialisation of these hierarchies.


In the current paper we turn this question around: 
The aim is to identify and classify those evolution 
equations/systems which can be multipotentialised into 
some given target potential equation/system.  This is the {\it converse problem}.
In principle, the converse problem consists of a
``{\it backwards-calculation-technique}'' that identifies 
both the equations and the potential 
variables that relates the equations to a given potential equation.
It is important to point out that the method proposed here does not
require the calculation of integrating factors for the 
equations/systems (see Proposition 1). 

\strut\hfill

To set the stage, we give an example of the {\it usual} (not {\it
  converse}) 
potentialisation
of a linear equation. Consider the following problem: 
Find all third-order evolution equations of the form
\begin{gather}
\label{F-v}
u_t=F(u,u_x,u_{xx},u_{xxx})
\end{gather}
that can be derived by the potentialisation of the linear equation
\begin{gather}
\label{E-eq}
E:=v_t-v_{xxx}=0.
\end{gather}
The corresponding auxiliary system for (\ref{E-eq}) is
\begin{subequations}
\begin{gather*}
u_x=\Phi^t(x,v,v_x,\ldots)\\
u_t=-\Phi^x(x,v,v_x,\ldots)
\end{gather*}
\end{subequations}
where
\begin{gather*}
\left.\vphantom{\frac{DA}{DB}}
D_t\Phi^t(x,v,\ldots)+D_x\Phi^x(x,v,\ldots)\right|_{v_t=v_{xxx}}=0.
\end{gather*}
Clearly $F$ in (\ref{F-v}) is not arbitrary but is constrained by
(\ref{E-eq}) and its corresponding $\Phi^t$ and $\Phi^x$.
In order to derive eq.(\ref{F-v}), we need to find all integrating factors,
$\Lambda(t,x,v,v_x,v_{xx},\ldots)$, for (\ref{E-eq}). Those can be
calculated by the conditions (see e.g. \cite{Euler_nonlocal_2009})
\begin{gather*}
\hat E[v]\left( \Lambda E\right)=0\ \ 
\Longleftrightarrow\ \ 
L_E^*[v]\Lambda=0,\ \ 
L_\Lambda[v]E=L^*_\Lambda[v]E,
\end{gather*}
where $\hat E[v]$ is the Euler operator
\begin{gather*}
\hat E[v]=\pde{\ }{v}-D_x\circ \pde{\ }{v_x}-D_t\circ\pde{\ }{v_t}+D^2_x\circ 
\pde{\ }{v_{2x}}-D_x^3\circ \pde{\ }{v_{3x}}+\cdots
\end{gather*}
and $L^*[v]$ is the adjoint of the linear operator $L[v]$,
\begin{gather*}
L[v]=\pde{E}{v}+\pde{E}{v_t}D_t+\pde{E}{v_x}D_x+\pde{E}{v_{xx}}D_x^2+\pde{E}{v_{xxx}}D_x^3,\\[0.3cm]
L^*[v]=\pde{E}{v}-D_t\circ\left(\pde{E}{v_t}\right)
-D_x\circ\left(\pde{E}{v_x}\right)
+D_x^2\circ \left(\pde{E}{v_{xx}}\right)
-D_x^3\circ \left(\pde{E}{v_{xxx}}\right).
\end{gather*}
The relation of $\Lambda$ to the conserved currents, $\Phi^t$, for (\ref{E-eq})
is 
\begin{gather*}
\Lambda=\hat E[v]\,  \Phi^t.
\end{gather*}
Following the above method, the only nonlinear equation of the form (\ref{F-v}), so
obtained, is \cite{Euler_NK_2009}
\begin{gather}
\label{v-3-4}
u_t=u_{xxx}-\frac{3}{4}\frac{u_{xx}^2}{u_x}.
\end{gather}

\begin{center}
\begin{displaymath}
\xymatrix{
\mbox{
{\bf Diagram 1: Potentialisation of $v_t=v_{xxx}$:}
}\\
\boxed{
\vphantom{\frac{DA}{DB}}
v_t=v_{xxx}
}
\ar[d]^{u_x=v^2}
\\
\boxed{
\vphantom{\frac{DA}{DB}}
u_t=u_{xxx}-\frac{3}{4}\frac{u_{xx}^2}{u_x}
}
}
\end{displaymath}
\end{center}

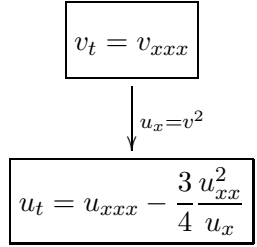
\strut\hfill

\noindent
In Section 3 we consider the converse problem of the above, i.e. we seek
the equations of the form (\ref{F-v})
for which (\ref{E-eq}) is the potential equation.
The results of the converse potentialisation are listed
as Case I in Section 3 and the results of the converse multipotentialisations of
(\ref{E-eq}) are listed in Case II and Case III (see Diagram 6).

\strut\hfill

The paper is organized as follows:
In Section 2 we give the main propositions that describes the
methodology of the proposed problem and introduce
triangular-auto-B\"acklund transformations. These transformations
act as solution generators for the equations.
In Sections 3 we classify
third-order evolution equations which can be
linearise by a suitable multipotentialisation.
For example, in this section we shown that the 
Calogero-Degasperis-Ibragimov-Shabat equation and the third-order
Burgers' equations, are just special cases of a class of 
third-order evolution equations which possess this type of 
linearisation property.
In Section 4 we study a fifth-order evolution equation and show 
that the converse multipotentialisation leads in a natural way to 
an interesting triangular-auto-B\"acklund transformation for the equation.
In Section 5 we propose the converse problem for systems 
of evolution in $(1+1)$ dimensions
and in Section 6 we extend our methodology to
evolution equations in higher dimensions. Some conluding remarks are
made in Section 7.

\section{The converse problem for the multipotentialisation of\\
  $(1+1)$-dimensional evolution equations}
In this section we consider $(1+1)$-dimensional evolution
equations and propose a method to study the converse problem
that aims to identify equations that can be potentialised in a target
potential equation. This addresses the problem of deriving
auto-B\"acklund transformations for evolution equations. 

\subsection{Definitions and Propositions}

Consider the following general $x$- and $t$-independent 
evolution equation of order $p$ in the form
\begin{gather}
\label{gen_eq1}
u_t=F(u,u_x,u_{xx},u_{3x},\ldots,u_{px}).
\end{gather}
We now define the converse problem and state conditions by which it can
be studied.

\strut\hfill

\noindent
{\bf Definition 1:} {\it
The {\bf converse problem} of the potentialisation of (\ref{gen_eq1})
aims to
determine the functional form(s) of $F$ in (\ref{gen_eq1})  for which 
(\ref{gen_eq1}) potentialises in a target equation of order $p$, given
by
\begin{gather}
\label{H-eq}
v_t=H(v_x,v_{xx},\ldots,v_{px})+\alpha_0 v,\qquad \alpha_0:\ \mbox{constant},
\end{gather}
with potential variable, $v$, and auxiliary system
\begin{subequations}
\begin{gather}
\label{1st_pot_a}
v_x=\Phi^t(x,u,u_x,\ldots)\\
\label{1st_pot_b}
v_t=-\Phi^x(x,u,u_x,\ldots),
\end{gather}
\end{subequations}
where
\begin{gather}
\label{conserv}
\left.
\vphantom{\frac{DA}{DB}}
D_t\Phi^t(x,u,u_x,\ldots)+D_x\Phi^x(x,u,u_x,\ldots)\right|_{u_t=F(u,u_x,u_{xx},\ldots,u_{px})}=0
\end{gather}
holds.
}

\strut\hfill

Following Definition 1 we replace $v_t$ from (\ref{H-eq}) in (\ref{1st_pot_b}), differentiate
(\ref{1st_pot_b})
with respect to $x$, and use (\ref{1st_pot_a}) and (\ref{conserv})
to express the resulting relation in terms of $\Phi^t$. This
leads to

\strut\hfill

\noindent
{\bf Proposition 1:}
{\it The condition on $\Phi^t$, such that 
\begin{gather*}
u_t=F(u,u_x,u_{xx},u_{3x},\ldots,u_{px}),
\end{gather*}
potentialises in 
\begin{gather*}
v_t=H(v_x,v_{xx},\ldots,v_{px})+\alpha_0 v,
\end{gather*}
is
\begin{gather}
\label{gen_cond_phi-t}
\left.
\vphantom{\frac{DA}{DB}}
D_xH\left(\Phi^t,\,D_x\Phi^t,D_x^2\Phi^t,\ldots,D_x^{p-1}\Phi^t\right)
+\alpha_0\Phi^t
=D_t\Phi^t\right|_{u_t=F(u,u_x,\ldots,u_{px})},
\end{gather}
where $H$ is a given function and $\alpha_0$ a given
constant.
}

\strut\hfill

\noindent
Note that condition (\ref{gen_cond_phi-t}) places a constrained on both
$\Phi^t$ and $F$ for a given $H$, which assures that (\ref{gen_eq1})
potentialises in (\ref{H-eq}).
Note that, in order to solve condition (\ref{gen_cond_phi-t})
for both $F$ and $\Phi^t$,
we need to make an assumption
regarding the functional dependence of $\Phi^t$. 
That is, we have to make a choice for 
the number of derivatives, $q$, allowed for $\Phi^t$:
\begin{gather*}
\Phi^t=\Phi^t(x,u,u_x,\ldots,u_{qx}).
\end{gather*}


Next we describe the {\it converse multipotentialisation process}.
Consider again the general equation, (\ref{gen_eq1}), viz.
\begin{gather*}
u_t=F(u,u_x,u_{xx},u_{3x},\ldots,u_{px}),
\end{gather*}
and assume that it can be potentialised in some
given evolution equation of order $p$, say
\begin{gather}
\label{eq_G}
v_t=G(v_x,v_{xx},v_{3x},\ldots,v_{px}),
\end{gather}
where (\ref{gen_eq1}) admits the auxiliary system
\begin{subequations}
\begin{gather}
\label{1st_pot_a_new}
v_x=\Phi_1^t(x,u,u_x,\ldots)\\
\label{1st_pot_b_new}
v_t=-\Phi_1^x(x,u,u_x,\ldots)
\end{gather}
\end{subequations}
and
\begin{gather}
\left.\vphantom{\frac{DA}{DB}}
D_t\Phi_1^t(x,u,\ldots)+D_x\Phi_1^x(x,u,\ldots)\right|_{u_t=F}=0.
\end{gather}
Introduce now a second auxiliary system, namely for (\ref{eq_G}),
of the form
\begin{subequations}
\begin{gather}
\label{Aux-w-1}
w_x=\Phi_2^t(x,v,v_x,\ldots)\\
\label{Aux-w-2}
w_t=-\Phi_2^x(x,v,v_x,\ldots),
\end{gather}
\end{subequations}
such that $w$ is the dependent variable for yet another evolution equation, say
\begin{gather}
\label{gen_H_w}
w_t=H(w_x,w_{xx},\ldots,w_{px})
\end{gather}
and
\begin{gather}
\left.\vphantom{\frac{DA}{DB}}
D_t\Phi_2^t(x,v,\ldots)+D_x\Phi_2^x(x,v,\ldots)
\right|_{v_t=G}=0.
\end{gather}
The above procedure provides a method to identify all equations of
the form (\ref{gen_eq1}) that can be 
potentialise in (\ref{eq_G})
under the first potential variable, $v$, with corresponding  
auxiliary system  (\ref{1st_pot_a_new})--(\ref{1st_pot_b_new}), and
which furthermore potentialises into (\ref{gen_H_w}) under the second
potential variable, $w$, with auxiliary system
(\ref{Aux-w-1})--(\ref{Aux-w-2}).
Hence this multipotentialisation procedure identifies 
the family of
equations, (\ref{gen_eq1}), that are related to (\ref{gen_H_w})
with a transformation that can be obtain by composing
\begin{subequations}
\begin{gather}
\label{Aux-v-1}
v_x=\Phi^t_1(x,u,u_x,\ldots)\\
\label{Aux-v-2}
w_x=\Phi^t_2(x,v,v_x,\ldots).
\end{gather}
\end{subequations}
We call this the {\it second-degree converse multipotentialisation} of (\ref{gen_H_w}).
The $n$th-degree converse multipotentialisations with potential variables,
$\{v_1,\ v_2,\ldots,v_{n-1},w\}$
can then be introduced  in an obvious manner, where
(\ref{Aux-v-1})--(\ref{Aux-v-2}) extends to
\begin{gather}
v_{1,x}=\Phi_1^t(x,u,u_x\ldots)\nn\\[0.3cm]
v_{2,x}=\Phi^t_2(x,v_1,v_{1,x},\ldots)\nn\\[0.3cm]
v_{3,x}=\Phi^t_3(x,v_2,v_{2,x},\ldots)\nn\\[0.3cm]
\label{gen-n-th-lin}
\qquad \vdots\\[0.3cm]
v_{(n-1),x}=\Phi^t_{n-1}(x,v_{n-2},v_{n-2,x},\ldots)\nn\\[0.3cm]
w_x=\Phi^t_n(x,v_{n-1},v_{n-1,x},\ldots).\nn
\end{gather}
Diagram 2 describes the $n$th degree converse multipotentialisation
of (\ref{gen_H_w}):
\begin{center}
\begin{displaymath}
\xymatrix{
\mbox{
{\bf Diagram 2:}
}\\
\mbox{
{\bf Converse multipotentialisation of $w_t=H$ of degree $n$}
}\\
\boxed{
\vphantom{\frac{DA}{DB}}
w_t=H(w_x,\ldots,w_{px})
}
\\
\boxed{
\vphantom{\frac{DA}{DB}}
v_{\{n-1\},t}=G_{n-1}(v_{\{n-1\},x},\, \ldots,\, v_{\{n-1\},px})
}
\ar[u]_{\ \ w_x=\Phi_n^t[v_{n-1}]}\\ 
\boxed{
\vphantom{\frac{DA}{DB}}
v_{2,t}=G_2(v_{2,x},\, \ldots,\, v_{2,px})
}
\ar@{.}[u] \\  
\boxed{
\vphantom{\frac{DA}{DB}}
v_{1,t}=G_1(v_{1,x},\, ,\ldots,\, v_{1,px})
}
\ar[u]_{\ \ v_{2,t}=\Phi_2^t[v_1]}\\
\boxed{
\vphantom{\frac{DA}{DB}}
u_t=F(u,u_x,\ldots,u_{px})
}
\ar[u]_{\ \ v_{1,t}=\Phi_1^t[u]}
}
\end{displaymath}
\end{center}



\subsection{Triangular auto-B\"acklund transformations}

In some cases we can combine and compose several conserved currents, $\Phi^t$,
to form nonpoint mappings of the dependent variable
of an equation to the same equation. This 
maps solutions to solutions and can hence be applied
to generate nontivial new solutions. We name such transformations
{\bf triangular Auto-B\"acklund transformation}, or 
{\bf $\bigtriangleup$-Auto-B\"acklund transformation}. There are
essentially three types of $\bigtriangleup$-Auto-B\"acklund
transformations. This is demonstrated in the Diagram 3 below. 
Note that ``Equation A $[V]$'' represents an evolution equation 
with $V$ as its dependent variable and $\Phi^t[V]$ denotes the equation's 
conserved current, which is a function of $x,V,\ V_x,\ V_{xx}$, etc.

{\tiny

\begin{displaymath}
\xymatrix{
&
\large{\mbox{{{{\bf Diagram 3}}}} }&\\
&
\large{\mbox{{{\bf$\bigtriangleup$-Auto-B\"acklund transformation: Type I }}} }&\\
& 
\boxed{
\vphantom{\frac{DA}{DB}}
\mbox{Equation B}\ [u]
}
&\\
 \boxed{
\vphantom{\frac{DA}{DB}}
\mbox{Equation A}\ [V]
}
\ar[ur]^{u_x=\Phi^t_2[V]\ \ } & 
\boxed{
\vphantom{\frac{DA}{DB}}
\ba{c}
\mbox{$\bigtriangleup$-Auto-B\"acklund}\\[0.2cm]
\Psi[v,V]:=\Phi_2^t[V]-\Phi_1^t[v]=0
\ea
}
\ar[l]&
\ar[ul]_{\ \ u_x=\Phi^t_1[v]}
\ar@{-}[l]
\boxed{
\vphantom{\frac{DA}{DB}}
\mbox{Equation A}\ [v]
}
\\
}
\end{displaymath}
}


{\tiny

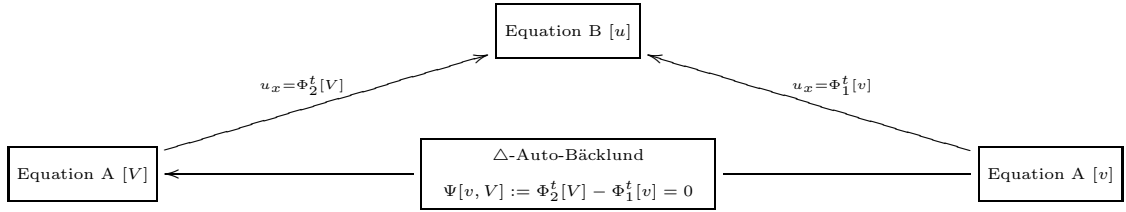
\begin{displaymath}
\xymatrix{
&
\large{\mbox{{{\bf$\bigtriangleup$-Auto-B\"acklund transformation: Type II }}} }&\\
& 
\boxed{
\vphantom{\frac{DA}{DB}}
\mbox{Equation B}\ [u]
}
\ar[dl]_{V_x=\Phi^t_2[u]\ \  }
&\\
 \boxed{
\vphantom{\frac{DA}{DB}}
\mbox{Equation A}\ [V]
}
& 
\boxed{
\vphantom{\frac{DA}{DB}}
\ba{c}
\mbox{$\bigtriangleup$-Auto-B\"acklund}\\[0.2cm]
\Psi[v,V]=0
\ea
}
\ar[l]&
\ar[ul]_{\ \ u_x=\Phi^t_1[v]}
\ar@{-}[l]
\boxed{
\vphantom{\frac{DA}{DB}}
\mbox{Equation A}\ [v]
}
\\
}
\end{displaymath}
}


{\tiny

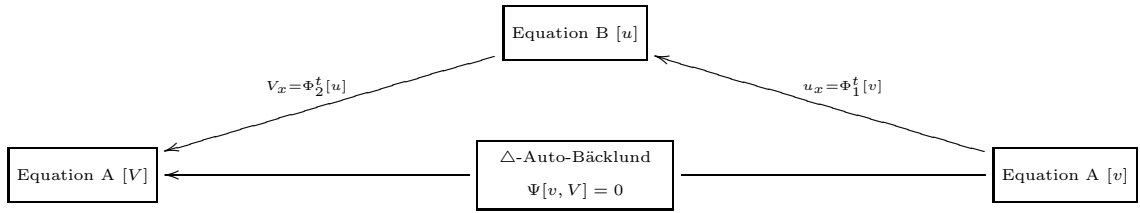
\begin{displaymath}
\xymatrix{
&
\large{\mbox{{{\bf$\bigtriangleup$-Auto-B\"acklund transformation: Type III }}} }&\\
& 
\boxed{
\vphantom{\frac{DA}{DB}}
\mbox{Equation B}\ [u]
}
\ar[dl]_{V_x=\Phi^t_2[u]\ \  }
\ar[dr]^{\ \ v_x=\Phi^t_1[u]}
&\\
 \boxed{
\vphantom{\frac{DA}{DB}}
\mbox{Equation A}\ [V]
}
& 
\boxed{
\vphantom{\frac{DA}{DB}}
\ba{c}
\mbox{$\bigtriangleup$-Auto-B\"acklund}\\[0.2cm]
\Psi[v,V]=0
\ea
}
\ar[l]&
\ar@{-}[l]
\boxed{
\vphantom{\frac{DA}{DB}}
\mbox{Equation A}\ [v]
}
\\
}
\end{displaymath}
}

Several $\bigtriangleup$-Auto-B\"acklund transformations are reported in
Propositions 2, 3, 4, 5 and Proposition 6.


\section{Third-order linearisable equations in $(1+1)$ dimensions} 

\subsection{First-degree converse potentialisation}

\noindent
For an application of Proposition 1, we now discuss the converse problem of
linesarisable evoluton equations, i.e. the problem by which to
determine the functional form(s) of $F$ in (\ref{gen_eq1}), {\it viz.}
\begin{gather*}
u_t=F(u,u_x,u_{xx},u_{3x},\ldots,u_{px}).
\end{gather*}
for which (\ref{gen_eq1})
potentialises in the linear evolution equation of order $p$,
\begin{gather}
\label{gen_lin}
v_t={\cal  L}^{(p)}[\alpha] v,
\end{gather}
under the first potential variable, $v$, with auxiliary system
(\ref{1st_pot_a}) - (\ref{1st_pot_b}). 
Here ${\cal L}^{(p)}$ is the general linear operator with parameters
$\{\alpha_0,\alpha_1,\ldots,\alpha_p\}$ defined by 
\begin{gather}
\label{lin-op}
{\cal L}^{(p)}[\alpha]:=\sum_{j=0}^p\alpha_j D_x^j.
\end{gather}
Note that
\begin{gather}
D_x\left.\vphantom{\frac{DA}{DB}}
{\cal L}^{(p)}[\alpha]\,v\right|_{v_x=\Phi^t}
={\cal L}^{(p)}[\alpha]\,\Phi^t.
\end{gather}
Following Proposition 1, the condition on $\Phi^t$ and $F$ for  potentialisation
the (\ref{gen_eq1}) in the linear equation (\ref{gen_lin}), then becomes
\begin{gather}
\label{main_cond}
\left.
\vphantom{\frac{A}{B}}
D_t\Phi^t\right|_{u_t=F}
={\cal L}^{(p)}[\alpha]\,\Phi^t.
\end{gather}



\noindent
As a special case we study third-order evolution equations with potentialisations in
\begin{gather}
\label{lin-eq-v}
v_t=v_{xxx}
\end{gather}
in detail. 
Consider the third-order evolution equations in the form
\begin{gather}
\label{auto_gen}
u_t=F(u,u_x,u_{xx},u_{xxx})
\end{gather}
and assume that (\ref{auto_gen}) admits a conserved current
of the form
\begin{gather}
\label{phi-gen}
\Phi^t=\Phi^t(u,u_x,u_{xx}).
\end{gather}
Solving condition (\ref{main_cond}), with the assumption of (\ref{phi-gen}),  
we find that  the most general form
of (\ref{auto_gen}) which
potentialises in the linear equation (\ref{lin-eq-v})
is given by the following two cases:

\strut\hfill

\noindent
{\bf Case I a:}
The conserved current
\begin{gather}
\label{phi-case2.2.a}
\Phi^t(u,u_x)=\frac{1}{\sqrt{2}}\left(\frac{u_x}{h}+c_1\right)^{1/2},   
\end{gather}
leads to the equation  
\begin{gather}
u_t=u_{xxx}
-\frac{3}{4}\left(\frac{u_{xx}^2}{u_x+c_1h}\right)
-\frac{3}{2}
\frac{h'}{h}
\left(\frac{u_x+2c_1h}{u_x+c_1h}\right)u_xu_{xx}
+\left(\frac{5}{4}\left(\frac{h'}{h}\right)^2-\frac{h''}{h}\right)u_x^3
\nonumber\\[0.3cm]
\label{eq-case2.2.a}
\qquad
+\frac{3}{4}\frac{c_1(h')^2}{h}\ u_x^2
-\frac{3}{4}c_1^2(h')^2\ u_x
-\frac{3}{4}\left(\frac{c_1^4h^2(h')^2}{u_x+c_1h}\right)
+\frac{3}{4}c_1^3(h')^2h+c_2h,
\end{gather}
where $h$ is an arbitrary but nonzero differentiable function of $u$ and $c_1$, $c_2$
are arbitrary constants.
\n\n
{\bf Case I b:}
The conserved current
\begin{gather}
\label{phi-case2.3}
\Phi^t(u,u_x,u_{xx})=\frac{1}{\sqrt{2hu_x+2c_1 h^2}}
\left(
u_{xx}-\left(\frac{h'}{h}\right)u_x^2
\right)
\end{gather}
leads to the equation  
\begin{gather}
u_t=u_{xxx}-\frac{3}{4}
\frac{u_{xx}^2}{u_x+c_1h}
-\frac{3}{2}\frac{h'}{h}
\left(\frac{u_x+2c_1h}{u_x+c_1h}\right)u_xu_{xx}
+\frac{1}{4}
\left(
\frac{5(h')^2-4hh''}{h^2(u_x+c_1h)}\right) u_x^4
\nn\\[0.3cm]
\label{eq-case2.3}
\qquad
+c_1\left(
\frac{2(h')^2-hh''}{h(u_x+c_1h)}\right)u_x^3
+c_2\left(
\frac{h}{u_x+c_1h}\right)u_x
+c_1c_2\left(\frac{h^2}{u_x+c_1h}\right),
\end{gather}
where $h$ is an arbitrary but nonzero differentiable function of $u$ and $c_1$, $c_2$
are arbitrary constants.

\strut\hfill

\noindent
{\bf Remark 1:}\n
The case, $\Phi^t=f_1(u)u_x+f_2(u)$ for any differentiable functions 
$f_1(u)$ and $f_2(u)$, result in linear equations for
(\ref{auto_gen}) under the point transformation $u\mapsto h(u)$ and are 
therefore not listed here.

\strut\hfill

\noindent
The above Case Ia and Case 1b lead to

\strut\hfill

\noindent
{\bf Proposition  2:}
{\it 
An $\bigtriangleup$-auto-B\"acklund transformation of type I for 
\begin{gather}
\label{eq-3-4}
u_t=u_{xxx}
-\frac{3}{4}\left(\frac{u_{xx}^2}{u_x}\right)
-\frac{3}{2}
\frac{h'}{h}u_xu_{xx}
+\left(\frac{5}{4}\left(\frac{h'}{h}\right)^2-\frac{h''}{h}\right)u_x^3
\end{gather}
is given by the relation 
\begin{gather}
\label{BT-1}
\frac{U_x}{h(U)}=\frac{1}{h(u)u_x}\,\left(u_{xx}-\left(\frac{h'(u)}{h(u)
}\right)\, u_x^2\right)^2,
\end{gather}
where $u$ and $U$ satisfy (\ref{eq-3-4})
for any nonzero arbitrary differentiable function $h$.
}

\strut\hfill

\noindent
{\bf Proof:}
Equations (\ref{eq-case2.2.a}) and (\ref{eq-case2.3}) with
\begin{gather}
\label{c012}
c_1=c_2=0 
\end{gather}
reduce to the same equation, namely (\ref{eq-3-4}). 
Consider now (\ref{eq-case2.2.a}) with (\ref{c012}) in terms of the dependent 
variable $U$, i.e.,
\begin{gather}
\label{eq-case2.2.a-s}
U_t=U_{xxx}
-\frac{3}{4}\left(\frac{U_{xx}^2}{U_x}\right)
-\frac{3}{2}
\frac{h'(U)}{h(U)}U_xU_{xx}
+\left(\frac{5}{4}\left(\frac{h'(U)}{h(U)}\right)^2-\frac{h''(U)}{h(U)}
\right)U_x^3
\end{gather}
with the conserved current, (\ref{phi-case2.3}), and its relation to
the potential variable $v$,
\begin{gather}
\label{vx-1}
v_x=\frac{1}{\sqrt{2}}\left(\frac{U_x}{h(U)}\right)^{1/2}.
\end{gather}
Moreover, (\ref{eq-3-4}) has 
the following relation to the same potential variable, $v$, namely
\begin{gather}
\label{vx-2}
v_x=\frac{1}{\sqrt{2h(u)u_x}}
\left(
u_{xx}-\left(\frac{h'(u)}{h(u)}\right)u_x^2.
\right)
\end{gather}
Relation (\ref{BT-1}), then 
follows by (\ref{vx-1}) and (\ref{vx-2}). 
\qquad $\boxed{\,}$

\strut\hfill

\noindent
{\bf Remark: 2:}
Equation (\ref{eq-3-4}) with $h(u)=1$, reduces to
\begin{gather}
\label{eq-3-4-sp}
u_t=u_{xxx}-\frac{3}{4}\frac{u_{xx}^2}{u_x}
\end{gather}
and the $\bigtriangleup$-auto-B\"acklund transformation, (\ref{BT-1}), takes the form
\begin{gather}
\label{AB-sp}
s_x=\frac{u_{xx}^2}{u_x}.
\end{gather}
This special case, (\ref{eq-3-4-sp}), and its auto-B\"acklund
transformation, (\ref{AB-sp}), has been reported in \cite{Euler_NK_2009}.


\subsection{Converse multipotentialisation}

For second degree converse multipotentialisations of the linear evolution equation 
\begin{gather}
\label{lin-gen-w}
w_t=w_{xxx}
\end{gather}
we consider (\ref{eq-3-4})
with 
\begin{gather}
h(u)=\exp(\alpha u),\qquad \alpha:\ \mbox{arbitrary constant},
\end{gather}
that is
\begin{gather}
\label{eq-3-4-alpha}
u_t=u_{xxx}-\frac{3}{4}\frac{u_{xx}^2}{u_x}-\frac{3}{2}\alpha
u_xu_{xx}+\frac{1}{4}\alpha^2u_x^3.
\end{gather}
We now construct the most general equation of the form
(\ref{auto_gen}), now written in terms of the variable $v$, 
\begin{gather}
\label{gen-p-eq}
v_t=F(v,v_x,v_{xx},v_{xxx}),
\end{gather}
which admits (\ref{eq-3-4-alpha}) as its potential equation 
with auxiliary system
\begin{subequations}
\begin{gather}
\label{a-p-1}
u_x=\Phi^t(v,v_x,v_{xx})\\
\label{a-p-2}
u_t=-\Phi^x(v,v_x,\ldots ).
\end{gather}
\end{subequations}
Applying Proposition 1 
we obtain the following constraint on $\Phi^t$:
\begin{gather}
D_x^3\Phi^t-\frac{3}{2}
\left(\Phi^t\right)^{-1}
D_x\Phi^t\, D_x^2\Phi^t
+\frac{3}{4}
\left(\Phi^t\right)^{-2}
\left(D_x\Phi^t\right)^3\nn\\[0.3cm]
\label{cond-2}
\left.
\vphantom{\frac{DA}{DB}}
\qquad 
-\frac{3\alpha}{2}
\left(D_x\Phi^t\right)^2
-\frac{3\alpha}{2}
\Phi^t D_x^2\Phi^t
+\frac{3\alpha^2}{4}
\left(
\Phi^t\right)^2
D_x \Phi^t
=D_t\Phi^t\right|_{v_t=F}.
\end{gather}

\strut\hfill

\noindent
By condition (\ref{cond-2}),
the most general form of (\ref{gen-p-eq}) for which
(\ref{eq-3-4-alpha})
is the potential form of (\ref{gen-p-eq}) with the conserved current
$\Phi^t=\Phi^t(v,v_x,v_{xx})$, is given by the following cases:\n\n
{\bf Case II a:} The conserved current
\begin{gather}
\Phi^t(v,v_x)=\frac{v_x}{f(v)}-c_1
\end{gather}
leads to the equation
\begin{gather}
\label{another-main-f}
v_t=v_{xxx}
+\frac{3}{4}\left(\frac{v_{xx}^2}{c_1f-v_x}\right)
-\frac{3}{2}
\left(
\frac{c_1f(\alpha+2f')-(f'+\alpha)v_x}{f(c_1f-v_x)}
\right)
v_xv_{xx}
+\frac{3}{2}\alpha c_1v_{xx}\nn\\[0.3cm]
\qquad
-\frac{1}{4}\left(
\frac{4f''f-5(f')^2-6\alpha f'-\alpha^2}{f^2}\right)v_x^3
-\frac{3}{4}\frac{c_1(f'+\alpha)^2}{f}\ v_x^2\nn\\[0.3cm]
\qquad
-\frac{3}{4}c_1^2\left((f')^2-\alpha^2\right) v_x
+\frac{3}{4}
\left(
\frac{(f'f)^2c_1^4}{c_1f-v_x}
\right)-\frac{3}{4}c_1^3f'f+c_2f,
\end{gather}
where $f$ is a nonzero arbitrary differentiable function of $v$ and $\alpha$, 
$c_1$, $c_2$ are arbitrary constants.
\n\n
{\bf Case II b:}
The conserved current
\begin{gather}
\Phi^t(v)=f(v),
\end{gather}
leads to the equation
\begin{gather}
v_t=v_{xxx}+\left(\frac{3f''}{f'}-\frac{3f'}{2f}\right)v_xv_{xx}
+\left(\frac{f'''}{f'}-\frac{3f''}{2f}+\frac{3}{4}\left(\frac{f'}{f}\right)^2\right)v_x^3
-\frac{3}{2}\alpha f v_{xx}\nn\\[0.3cm]
\label{main-f}
\qquad +\frac{3}{4}\alpha^2f^2v_x
-\frac{3}{2}\alpha 
\left(f'+\frac{ff''}{f'}\right)v_x^2,
\end{gather}
where $f$ is a nonconstant arbitrary differentiable function of $v$ and $\alpha$
is an arbitrary constant.
\n\n
{\bf Case II c:}
For $\alpha=0$, the conserved current
\begin{gather}
\Phi^t(v)=\frac{(f'(v))^2}{f(v)}v_x^2,\qquad f'(v)\neq 0,
\end{gather}
leads to the equation
\begin{gather}
\label{ab-eq-f}
v_t=v_{xxx}+\left(\frac{3f''}{f'}-\frac{3f'}{2f}\right)v_xv_{xx}
+\left(\frac{f'''}{f'}-\frac{3f''}{2f}+\frac{3}{4}\left(\frac{f'}{f}\right)^2\right)v_x^3,
\end{gather}
where $f$ is a nonconstant arbitrary differentiable function of $v$.

\strut\hfill

\noindent
{\bf Remark 3:}\n
It is interesting to note that (\ref{main-f}) contains, for
special values of $\alpha$ and special functions $f$, two
well-known equations, namely the following:

\strut\hfill

\noindent
With $\alpha =-2$ and $f(v)=v^2$ equation (\ref{main-f}) is the
Calogero-Degasperis-Ibragimov-Shabat equation (CDIS)
(\cite{Calogero87}, \cite{Ibragimov_Shabat})
\begin{gather}
\label{cdis}
v_t=v_{xxx}+3v^2v_{xx}+9vv_x^2+3v^4v_x
\end{gather}
and with $\alpha =0$ and $f(v)=\exp(2v)$ equation (\ref{main-f}) is the
third-order potential Burgers' equation \cite{Euler_nonlocal_2009}
\begin{gather}
\label{burgers}
v_t=v_{xxx}+3v_xv_{xx}+v_x^3.
\end{gather}
In \cite{Euler_nonlocal_2009} we showed that
both (\ref{cdis}) and (\ref{burgers}) linearise under a
suitable multipotentialisation.
Hence the equation (\ref{main-f}) can be viewed as a generalisation of the
Calogero-Degasperis-Ibragimov-Shabat equation, (\ref{cdis}),
and the third-order Burgers' equation, (\ref{burgers}), as
(\ref{main-f}) combines both of these interesting equations into
a single equation with arbitrary function, $f(v)$. See also Diagram 6. 
\begin{center}
\begin{displaymath}
\xymatrix{
\mbox{
{\bf Diagram 4: Converse potentialisation of $w_t=w_{xxx}$:}
}\\
\boxed{
\vphantom{\frac{DA}{DB}}
w_t=w_{xxx}
}
\\
\boxed{
\vphantom{\frac{DA}{DB}}
\mbox{{\bf Case II}} [v]
\ar[u]_{w_x=\Phi^t[v]}
}
}
\end{displaymath}
\end{center}

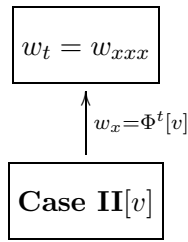
\strut\hfill

\noindent
A closer look ar Case IIb and Case IIc reveals a 
$\bigtriangleup$-auto-B\"acklund transformation for (\ref{ab-eq-f}). 

\strut\hfill

\noindent
{\bf Proposition 3:} 
{\it
An $\bigtriangleup$-auto-B\"acklund transformation of type I
for (\ref{ab-eq-f}), viz.
\begin{gather*}
v_t=v_{xxx}+\left(\frac{3f''}{f'}-\frac{3f'}{2f}\right)v_xv_{xx}
+\left(\frac{f'''}{f'}-\frac{3f''}{2f}+\frac{3}{4}\left(\frac{f'}{f}\right)^2\right)v_x^3,
\end{gather*}
is given by the relation
\begin{gather}
f(V)=\frac{\left(f'(v)\right)^2}{f(v)}\ v_x^2,
\end{gather}
where $v$ and $V$ satisfy (\ref{ab-eq-f})
for any nonconstant differentiable function $f$}

\strut\hfill

\noindent
Applying Proposition 3 with $f(v)=e^{2v}$ on the third-order
potential  Burgers' equation, 
(\ref{burgers}), {\it viz.}
\begin{gather*}
v_t=v_{xxx}+3v_xv_{xx}+v_x^3,
\end{gather*}
we obtain the $\bigtriangleup$-auto-B\"acklund
transformation of type I for (\ref{burgers}) in
the form
\begin{gather}
\label{AB1-Burgers}
e^{2V}=4e^{2v}v_x^2.
\end{gather}
By differentiating (\ref{AB1-Burgers}) we arrive at the relation
\begin{gather}
\label{AB-Burgers}
V_x=v_x+D_x \ln|v_x|
\end{gather}
which can be applied to gain auto-B\"acklund transformations for
those equations which can be potentialised in (\ref{burgers}).  
Note that (\ref{eq-3-4-alpha})
with $\alpha=0$, i.e.,
\begin{gather}
\label{eq3-4-a0}
u_t=u_{xxx}-\frac{3}{4}\frac{u_{xx}^2}{u_x},
\end{gather}
and (\ref{burgers}), both admit linear integro-differential
recursion operators: Equation (\ref{eq3-4-a0}) admits the
second-order recursion operator, $R_1[u]$, given by (e.g. \cite{EE-RO-1})
\begin{gather}
R_1[u]=D_x^2-\frac{u_{xx}}{u_x}D_x
+\frac{1}{2}\frac{u_{xx}}{u_x}
-\frac{1}{4}\left(\frac{u_{xx}}{u_x}\right)^2\\[0.3cm]
\qquad
-\frac{1}{2}D_x^{-1}\circ \left(
\frac{u_{xxxx}}{u_x}-\frac{2u_{xx}u_{xxx}}{u_x^2}
+\left(\frac{u_{xx}}{u_x}\right)^3\right),
\end{gather}
whereas (\ref{burgers}) admits the first-order recursion operator,
$R_2[v]$, given by (e.g. \cite{EEP})
\begin{gather}
R_2[v]=D_x+v_x.
\end{gather}
Equations (\ref{eq3-4-a0}) and (\ref{AB-Burgers})
can now be written, respectively, in the form
\begin{gather}
u_t=R_1[u]u_x,\qquad v_t=R_2^2[v] v_x,
\end{gather}
and the hierarchies of $n$ equations are
\begin{subequations}
\begin{gather}
\label{H-1}
u_t=R_1^n[u] u_x\\
\label{H-2}
v_t=R_2^n[v] v_x,\qquad n\in {\cal N}.
\end{gather}
\end{subequations}
Hierarchy (\ref{H-2}) is known as the potential Burger' hierarchy 
\cite{Euler_nonlocal_2009}.
Since all equations in a given hierarchy of evolution equations admit
the same conserved currents, the $\bigtriangleup$-auto-B\"acklund transformation
for (\ref{AB-Burgers}) is valid for the entire potential 
Burgers' hierarchy (\ref{H-2}). The transformations between
the two hierarchies and their $\bigtriangleup$-auto-B\"acklund
transformation are illustrated in Diagram 5.



\begin{center}
\begin{displaymath}
\xymatrix{
&
\mbox{{{\bf Diagram 5}}} &\\
& 
\boxed{
\vphantom{\frac{DA}{DB}}
u_t=R_1^n[u]u_x}
&\\
 \boxed{
\vphantom{\frac{DA}{DB}}
V_t=R^{2n}_2[V]V_x}
\ar[ur]^{u_x=\exp(2V)\ \ } & 
\boxed{
\vphantom{\frac{DA}{DB}}
V_x=v_x+D_x\ln|v_x|}
\ar[l]&
\ar[ul]_{\ \ u_x=4v_x^2\exp(2v)}\ar@{-}[l]
\boxed{
\vphantom{\frac{DA}{DB}}
v_t=R^{2n}_2[v]v_x}
\\
}
\end{displaymath}
\end{center}

\strut\hfill

We now consider (\ref{burgers}), {\it viz.}
\begin{gather*}
v_t=v_{xxx}+3v_xv_{xx}+v_x^3,
\end{gather*}
for the third degree converse potentialisation
of (\ref{lin-gen-w}).
Applying Proposition 1, we obtain the constraint 
\begin{gather}
\label{cont-burgers}
\left.
\vphantom{\frac{DA}{DB}}
D_x^3\Phi^t+3\left(D_x\Phi^t\right)^2
+3\Phi^t D_x^2\Phi^t
+3\left(\Phi^t\right)^2 D_x\Phi^t=D_t\Phi^t\right|_{q_t=F(q,q_x,q_{xx},q_{xxx})},
\end{gather}
which allows 
\begin{gather}
q_t=F(q,q_x,q_{xx},q_{xxx})
\end{gather}
to be potentialised in (\ref{burgers}) with the auxiliary system
\begin{subequations}
\begin{gather}
v_x=\Phi^t(q,q_x)\\
v_t=-\Phi^x(q,q_x.\ldots).
\end{gather}
\end{subequations}
This identifies five cases:

\strut\hfill

\noindent
{\bf Case III a:} The conserved current
\begin{gather}
\Phi^t(q)=g(q),\qquad g'(q)\neq 0
\end{gather}
leads to the equation
\begin{gather}
\label{gen-burg}
q_t=q_{xxx}+3\left(\frac{g''}{g'}\right)q_xq_{xx}+3gq_{xx}
+\left(\frac{g'''}{g'}\right)q_x^3+3\left(g'+\frac{gg''}{g'}\right)
q_x^2+3g^2 q_x,
\end{gather}
where $g$ is an arbitrary nonconstant differentiable function of
$q$.\n\n
{\bf Note:} With $g=q$, (\ref{gen-burg}) is the third-order
Burgers' equation \cite{EEP}, \cite{EE}, 
\begin{gather}
q_t=q_{xxx}+3q_x^2+3qq_{xx}+3q^2q_x.
\end{gather}

\strut\hfill

\noindent
{\bf Case III b:}
The conserved current
\begin{gather}
\Phi^t(q,q_x)=g(q)q_x+c_1,\qquad g(q)\neq 0
\end{gather}
leads to the equation
\begin{gather}
q_t=q_{xxx}
+3\left(\frac{g'}{g}+g\right)q_xq_{xx}
+2c_1q_{xx}
+\left(3g'+g^2+\frac{g''}{g}\right) q_x^3\nn\\[0.3cm]
\label{caseIIIb}
\qquad
+3c_1\left(g+\frac{g'}{g}\right)q_x^2
+3c_1^2 q_x+\frac{c_2}{g},
\end{gather}
where $g$ is an arbitrary nonzero differentiable function of $q$
and $c_1,\ c_2$ are arbitrary constants.

\strut\hfill

\noindent
{\bf Case III c:}
The conserved current
\begin{gather}
\Phi^t(q,q_x)=\left(\frac{g'(q)}{g(q)+c_1}\right)q_x+g(q),\qquad g'(q)\neq 0
\end{gather}
leads to the equation
\begin{gather}
q_t=q_{xxx}+3\left(\frac{gg''+c_1g''+(g')^2}{g'(g+c_1)}\right) q_xq_{xx}
+3\left(\frac{g'}{g+c_1}\right) q_xq_{xx}
+3gq_{xx}\nn\\[0.3cm]
\qquad
+\left(\frac{g'''}{g'}\right)q_x^3
+3\left(\frac{gg''}{g'}+g'\right) q_x^2
+3g^2 q_x,
\end{gather}
where $g$ is an arbitrary nonconstant differentiable function of $q$
and $c_1$ is an arbitrary constant.

\strut\hfill

\noindent
{\bf Case III d:}
The conserved current
\begin{gather}
\Phi^t(q,q_x)=\left(\frac{1}{2}\frac{g'(q)}{g(q)}\right)\ q_x+g(q),\qquad g'(q)\neq 0
\end{gather}
leads to the equation
\begin{gather}
q_t=q_{xxx}
+\frac{3}{2}\left(
\frac{2gg''-(g')^2}{gg'}\right)\ q_xq_{xx}
+3gq_{xx}\nn\\[0.3cm]
\qquad
+\frac{1}{4}\left(\frac{3(g')^3-6gg'g''+4g^2g'''}{g^2g'}\right)\ q_x^3
+3\left(g'+\frac{gg''}{g'}\right)\ q_x^2+3g^2q_x,
\end{gather}
where $g$ is an arbitrary nonconstant differentiable function of $q$.

\strut\hfill

\noindent
{\bf Case III e:}
The conserved current
\begin{gather}
\Phi^t(q,q_x)=\sqrt{Q}+g(q),
\end{gather}
where
\begin{gather}
Q:=g'q_x+g^2+c_1,\qquad g'(q)\neq 0,
\end{gather}
leads to the equation
\begin{gather}
q_t=q_{xxx}+\left(\frac{g'}{Q}\right)\ q_{xx}^2
+\frac{3}{2}\left(\frac{g''}{Q}\right)q_x^2q_{xx}
+\left(\frac{3}{Qg'}\right)\left(
g^2g''+c_1g''-\sqrt{Q}(g')^2
\right)q_xq_{xx}\nn\\[0.3cm]
\qquad
+\frac{3}{Q}\left(
g^3+c_1g-\sqrt{Q}g^2-c_1\sqrt{Q}
\right)q_{xx}
+\frac{1}{Qg'}\left(
g'g'''-\frac{3}{4}(g'')^2\right)q_x^4\nn\\[0.3cm]
\qquad
+\frac{1}{Qg'}
\left(
g^2g'''+c_1g'''+6(g')^3-3\sqrt{Q}g'g''
\right)q_x^3\nn\\[0.3cm]
\qquad
+\frac{3}{Qg'}\left[
gg''\left(c_1-\sqrt{Q}g+g^2\right)
+(g')^2\left(
3g^2
+3c_1-2\sqrt{Q}g\right)
-c_1\sqrt{Q}g''\right] q_x^2\nn\\[0.3cm]
\qquad
+\frac{3}{Q}\left(
c_1^2-2c_1\sqrt{Q}g+3c_1g^2-2\sqrt{Q}g^3+2g^4
\right) q_x.
\end{gather}
Here $g$ is an arbitrary nonconstant differentiable function of $q$
and $c_1$ is an arbitrary constant.

\strut\hfill

\noindent
A detailed graphical description of the converse multipotentialisation of (\ref{lin-gen-w})
is given in Diagram 6.

\strut\hfill

As described in Section 2, the linearisation transformations of all
the equations listed above can now be determined by composing the
corresponding conserved currents. For example, eq. (\ref{main-f}) of 
Case II b linearises in (\ref{lin-gen-w})
under the nonlocal transformation
\begin{gather}
w_x=\frac{1}{\sqrt{2}}f(v)^{1/2}\exp\left(
-\frac{\alpha}{2}\int f(v)\,dx\right),
\end{gather}
which is obtained by composing
\begin{subequations}
\begin{gather}
w_x=\frac{1}{\sqrt{2}}\,u_x^{1/2}\exp\left(-\frac{\alpha}{2}u\right)\\
u_x=f(v).
\end{gather}
\end{subequations}

{\tiny
\begin{landscape}
\begin{center}
\begin{displaymath}
\xymatrix{
&\mbox{\large{
{\bf Diagram 6:}}}
\\
\boxed{
\ba{l}
u_t=u_{xxx}-\frac{3}{4}
\frac{u_{xx}^2}{u_x+c_1h}
-\frac{3}{2}\frac{h'}{h}
\left(\frac{u_x+2c_1h}{u_x+c_1h}\right)u_xu_{xx}
+\frac{1}{4}
\left(
\frac{5(h')^2-4hh''}{h^2(u_x+c_1h)}\right) u_x^4\\[0.3cm]
\qquad +c_1\left(
\frac{2(h')^2-hh''}{h(u_x+c_1h)}\right)u_x^3
+c_2\left(
\frac{h}{u_x+c_1h}\right)u_x
+c_1c_2\left(\frac{h^2}{u_x+c_1h}\right)
\ea
}
\ar[r]^{\qquad\qquad\qquad\qquad\qquad\qquad\qquad
w_x=\frac{1}{\sqrt{2hu_x+2c_1 h^2}} 
\left(
u_{xx}-\left(\frac{h'}{h}\right)u_x^2\right)
}
&
\boxed{
\vphantom{\frac{DA}{DB}}
w_t=w_{xxx}
}
\\
& 
\boxed{
\ba{l}
u_t=u_{xxx}
-\frac{3}{4}\left(\frac{u_{xx}^2}{u_x+c_1h}\right)
-\frac{3}{2}
\frac{h'}{h}
\left(\frac{u_x+2c_1h}{u_x+c_1h}\right)u_xu_{xx}
+\left(\frac{5}{4}\left(\frac{h'}{h}\right)^2-\frac{h''}{h}\right)u_x^3\\[0.3cm]
\qquad +\frac{3}{4}\frac{c_1(h')^2}{h}\ u_x^2
-\frac{3}{4}c_1^2(h')^2\ u_x
-\frac{3}{4}\left(\frac{c_1^4h^2(h')^2}{u_x+c_1h}\right)
+\frac{3}{4}c_1^3(h')^2h+c_0h
\ea
}
\ar[u]_{w_x=\frac{1}{\sqrt{2}}\sqrt{c_1+u_x/h(u)}}
\\ 
\boxed{
\ba{l}
v_t=v_{xxx}
+\frac{3}{4}\left(\frac{v_{xx}^2}{c_1f-v_x}\right)
-\frac{3}{2}
\left(
\frac{c_1f(\alpha+2f')-(f'+\alpha)v_x}{f(c_1f-v_x)}
\right)
v_xv_{xx}
+\frac{3}{2}\alpha c_1v_{xx}\nn\\[0.3cm]
\qquad
-\frac{1}{4}\left(
\frac{4f''f-5(f')^2-6\alpha f'-\alpha^2}{f^2}\right)v_x^3
-\frac{3}{4}\frac{c_1(f'+\alpha)^2}{f}\ v_x^2\\[0.3cm]
\qquad
-\frac{3}{4}c_1^2\left((f')^2-\alpha^2\right) v_x
+\frac{3}{4}
\left(
\frac{(f'f)^2c_1^4}{c_1f-v_x}
\right)-\frac{3}{4}c_1^3f'f+c_2f
\ea
}
\ar[r]^{\qquad\qquad u_x=v_x/f-c_1}& 
\boxed{
u_t=u_{xxx}-\frac{3}{4}\frac{u_{xx}^2}{u_x}-\frac{3}{2}\alpha
u_xu_{xx} +\frac{\alpha^2}{4}u_x^3
}
\ar@{=}[u]_{\quad h(u)=\exp(\alpha u)}^{\qquad c_1=0\quad }
\\ 
&  \ar[u]_{u_x=f(v)} 
\boxed{
\ba{l}
v_t=v_{xxx}+\left(\frac{3f''}{f'}-\frac{3f'}{2f}\right)v_xv_{xx}
+\left(\frac{f'''}{f'}-\frac{3f''}{2f}+\frac{3}{4}\left(\frac{f'}{f}\right)^2\right)v_x^3
-\frac{3}{2}\alpha f v_{xx}\\[0.2cm]
\qquad +\frac{3}{4}\alpha^2f^2v_x
-\frac{3}{2}\alpha 
\left(f'+\frac{ff''}{f'}\right)v_x^2
\ea
}
\\ \ar@{=}[ur]^{f(v)=v^2\ \qquad \qquad}_{\quad \alpha=-2}
\mbox{CDIS-equation}\quad
\boxed{
\vphantom{\frac{DA}{DB}}
v_t=v_{xxx}+3v^2v_{xx}+9vv_x^2+3v^4v_x
}
& 
\qquad\qquad\qquad\qquad\qquad\qquad\qquad\quad
\boxed{
\vphantom{\frac{DA}{DB}}
v_t=v_{xxx}+3v_xv_{xx}+v_x^3
}\quad \mbox{3rd-order potential Burgers' eq.}
\ar@{=}[u]^{f(v)=\exp(2v)\ \ }_{\quad \alpha=0}
\\ & \ar[u]_{\Phi^t(q,q_x)} 
\boxed{
\mbox{\vphantom{$\frac{A}{B}$}
{\bf Case III a, Case III b, Case III c, Case III d, Case III e}}
}
}
\end{displaymath}
\end{center}
\end{landscape}
}

\noindent
By Proposition 3 and the auto-B\"acklund transformation
(\ref{AB-Burgers}) for the potential Burgers equation (\ref{burgers}),
we have auto-B\"acklund transformations
for all equations listed in Case III above (see Diagram 7). 
These auto-B\"acklund transformation are of the form
\begin{gather}
\Phi^t(Q,Q_x)= 
\Phi^t(q,q_x)+D_x \ln|\Phi^t(q,q_x)|,
\end{gather}
where $\Phi^t$ are the conserved currents of the equations in Case
III. For example, the equation given in Case IIIa, namely
(\ref{gen-burg}),
\begin{gather*}
q_t=q_{xxx}+3\left(\frac{g''}{g'}\right)q_xq_{xx}+3gq_{xx}
+\left(\frac{g'''}{g'}\right)q_x^3+3\left(g'+\frac{gg''}{g'}\right)
q_x^2+3g^2 q_x,
\end{gather*}
admits the auto-B\"acklund transformation
\begin{gather}
\label{AB-B}
g(Q)=g(q)+\frac{g'(q)}{g(q)}\ q_x
\end{gather}
where $\Phi^t=g(q)$. For $g(q)=q$, (\ref{gen-burg}) is the well-known
third-order Burgers' equation \cite{Euler_nonlocal_2009}
\begin{gather}
q_t=q_{xxx}+3qq_{xx}+3q_x^2+3q^2 q_x,
\end{gather}
and (\ref{AB-B}) reduces to the well-known auto-B\"acklund
transformation
\begin{gather}
Q=q+\frac{q_x}{q}
\end{gather}
which can be derived by a truncated Painlev\'e expansion for the
Burgers' equation (see e.g. \cite{SE}).
As a second example, consider Case III b. It follows that
\begin{gather}
g(Q)Q_x=g(q)q_x
+\frac{g'(q)q_x^2+g(q)q_{xx}}{g(q) q_x+c_1}
\end{gather}
is an auto-B\"acklund transformation for (\ref{caseIIIb}).

{\tiny
\begin{center}

\begin{displaymath}
\xymatrix{
& \mbox{{\large {\bf Diagram 7: }}} &\\
& 
\boxed{
\vphantom{\frac{DA}{DB}}
u_t=u_{xxx}-\frac{3}{4}\frac{u_{xx}^2}{u_x} } 
&\\
\boxed{
\vphantom{\frac{DA}{DB}}
V_t=V_{xxx}+3V_xV_{xx}+V_x^3 } 
\ar[ur]^{u_x=\exp(2V)\ \ } & 
\boxed{
\vphantom{\frac{DA}{DB}}
V_x=v_x+D_x\ln|v_x| }
\ar[l]&
\ar[ul]_{\ \ \ u_x=4v_x^2\exp(2v)} 
\ar@{-}[l] 
\boxed{
\vphantom{\frac{DA}{DB}}
v_t=v_{xxx}+3v_xv_{xx}+v_x^3 } 
\\
\boxed{
\vphantom{\frac{DA}{DB}}
\ba{l}
\mbox{{\bf Case III: a), b), c), d), e)} }\\[0.2cm]
Q_t=F(Q,Q_x,Q_{xx},Q_{xxx})
\ea 
} 
\ar[u]^{V_x=\Phi^t(Q,Q_x)} & 
\boxed{
\vphantom{\frac{DA}{DB}}
\Phi^t(Q,Q_x)= 
\Phi^t(q,q_x)+D_x\ln|\Phi^t(q,q_x)|
}
\ar[l]& \ar@{-}[l] 
\boxed{
\vphantom{\frac{DA}{DB}}
\ba{l}
\mbox{{\bf Case III: a), b), c), d), e)} }\\[0.2cm]
q_t=F(q,q_x,q_{xx},q_{xxx})
\ea 
} 
\ar[u]_{v_x=\Phi^t(q,q_x)}\\
}
\end{displaymath}
\end{center}
}

\section{Converse multipotentialisation of a 
fifth-order integrable evolution equation}

In this section we apply the converse
multipotentialisation methodology
on the following fifth-order equation:
\begin{gather}
\label{Reyes-u}
u_t=u_{5x}-\frac{5u_{xx}u_{4x}}{u_x}
-\frac{15}{4}\frac{u_{xxx}^2}{u_x}
+\frac{65}{4}\frac{u_{xx}^2u_{xxx}}{u_x^2}-\frac{135}{16}\frac{u_{xx}^4}{u_x^3}.
\end{gather}
Equation (\ref{Reyes-u}) plays a central role in the nonlocal
invariance of the Kaup-Kupershmidt equation \cite{Reyes}.
We show that a converse multipotentialisation of (\ref{Reyes-u})
leads to a $\bigtriangleup$-auto-B\"acklund transformation of type II.

Our aim is to find 5th-order equations of the form
\begin{gather}
\label{5th-v}
v_t=F(v,v_x,v_{xx},\ldots,v_{5x}),
\end{gather}
such that (\ref{5th-v}) potentialises in (\ref{Reyes-u}). The auxiliary
system for (\ref{5th-v}) is
\begin{subequations}
\begin{gather}
u_x=\Phi_1^t(v,v_x,\ldots)\\
u_t=-\Phi_1^x(v,v_x,\ldots).
\end{gather}
\end{subequations}
By Proposition 1 we obtain the following condition on $\Phi_1^t$:
\begin{gather}
D_x^5 \Phi_1^t
-\frac{25}{2} \left(\Phi_1^t\right)^{-1} 
D_x^2\Phi^t D_x^3\Phi^t
-5\left(\Phi_1^t\right)^{-1} 
D_x \Phi_1^t D_x^4\Phi_1^t
+\frac{85}{4}\left(\Phi_1^t\right)^{-2}
\left(D_x \Phi_1^t\right)^2 D_x^3\Phi_1^t\nn\\[0.3cm]
\qquad +\frac{145}{4}\left(\Phi_1^t\right)^{-2}
D_x \Phi_1^t \left(D_x^2\Phi_1^t\right)^2
-\frac{265}{4}\left(\Phi_1^t\right)^{-3}
\left(D_x \Phi_1^t\right)^3 D_x^2\Phi_1^t\nn\\[0.3cm]
\label{5th-Phi-1}
\qquad
\left.\vphantom{\frac{DA}{DB}}
+\frac{405}{16}\left(\Phi_1^t\right)^{-4}
\left(D_x \Phi_1^t\right)^5
=D_t \Phi_1^t\right|_{v_t=F(v,v_x,\ldots,v_{5x})}.
\end{gather}
A first-degree converse potentialisation of (\ref{Reyes-u})
is then obtained by solving (\ref{5th-Phi-1})
for $\Phi_1^t$ and $F$. One of the solutions is
\begin{gather}
\label{5th-CC-v}
\Phi_1^t=v^4v_x^{-2}
\end{gather}
for the equation
\begin{gather}
\label{5th-v-eq}
v_t=v_{5x}-\frac{5v_{xx}v_{4x}}{v_x}
+\frac{5v_{xx}v_{xxx}}{v_x^2}
\end{gather}
For a second-degree converse multipotentialisation of (\ref{Reyes-u}),
we apply a first-degree converse potentialisation on (\ref{5th-v-eq}).
That is, we seek an equation of the form
\begin{gather}
\label{5th-CC-V}
V_t=G(V,V_x,V_{xx},\ldots,V_{5x})
\end{gather}
that would potentialise in (\ref{5th-v-eq}). 
The auxiliary system for (\ref{5th-CC-V})
is
\begin{subequations}
\begin{gather}
v_x=\Phi_2^t(V,V_x,\ldots)\\
v_t=-\Phi_2^x(V,V_x,\ldots)
\end{gather}
\end{subequations}
and by Proposition 1 we obtain the following condition on $\Phi_2^t$:
\begin{gather}
D_x^5\Phi_2^t
+5\left(\Phi_2^t\right)^{-2}\left(D_x\Phi_2^t\right)^2D_x^3\Phi_2^t
-5\left(\Phi_2^t\right)^{-1}
D_x^2\Phi_2^tD_x^3\Phi^t
-5\left(\Phi_2^t\right)^{-1}
D_x\Phi_2^tD_x^4\Phi^t\nn\\[0.3cm]
\qquad
-10\left(\Phi_2^t\right)^{-3}
\left(D_x\Phi_2^t\right)^2 D_x^2\Phi_2^t
+5\left(\Phi_2^t\right)^{-2}
\left(D_x^2\Phi_2^t\right)^2 \nn\\[0.3cm]
\label{5th-Phi-2}
\qquad
\left.\vphantom{\frac{DA}{DB}}
+5\left(\Phi_2^t\right)^{-2}
D_x\Phi_2^t D_x^3\Phi_2^t
=D_t\Phi_2^t\right|_{V_t=G(V,V_x,\ldots,V_{5x})}.
\end{gather}
A solution of (\ref{5th-Phi-2})
is
\begin{gather}
\label{5th-V}
\Phi_2^t=V\,V_x^{-1/2},
\end{gather}
for the equation
\begin{gather}
\label{5th-V-Eq}
V_t=
V_{5x}-\frac{5V_{xx}V_{4x}}{V_x}
-\frac{15}{4}\frac{V_{xxx}^2}{V_x}
+\frac{65}{4}\frac{V_{xx}^2V_{xxx}}{V_x^2}
-\frac{135}{16}\frac{V_{xx}^4}{V_x^3}.
\end{gather}
We note in passing that the equations (\ref{5th-V-Eq})
and (\ref{Reyes-u}) are identical equations. Hence we have 
a $\bigtriangleup$-auto-B\"acklund transformation 
of type II for 
equation (\ref{Reyes-u}). This auto-B\"acklund transformation is 
given by the composition of
\begin{gather}
u_x=v^4v_x^{-1},\qquad v_x=V\,V_x^{-1/2}
\end{gather}
that leads to

\strut\hfill

\noindent
{\bf Proposition 4:} {\it A $\bigtriangleup$-auto-B\"acklund
transformation of type II
for (\ref{Reyes-u}), viz.
\begin{gather*}
u_t=u_{5x}-\frac{5u_{xx}u_{4x}}{u_x}
-\frac{15}{4}\frac{u_{xxx}^2}{u_x}
+\frac{65}{4}\frac{u_{xx}^2u_{xxx}}{u_x^2}-\frac{135}{16}\frac{u_{xx}^4}{u_x^3},
\end{gather*}
is given by the relation
\begin{gather}
W_x=\frac{1}{4}\left(\frac{V_{xx}}{V_x}-\frac{2V_x}{V}\right)
W+\frac{V^{1/2}}{V_x^{1/4}}
\end{gather}
with 
\begin{gather}
W^4(x,t)=\pde{u(x,t)}{x},
\end{gather}
where $V$ and $u$ satisfy equation (\ref{Reyes-u}).
}

 \section{Systems of evolution equations in $(1+1)$ dimensions}
We consider a system of $m$ evolution equations of order $p$ in the form
\begin{gather}
\label{sys-F-j}
u_{j,t}=F_j({\bf u},\,{\bf u}_x,\ldots,{\bf u}_{px}),\qquad j=1,2,\ldots,m, 
\end{gather}
where
\begin{gather*}
{\bf u}:=(u_1,u_2,\ldots,u_m),\qquad
{\bf u}_x:=(u_{1,x},u_{2,x},\ldots,u_{m,x}),\ \ldots,\\[0.3cm]
{\bf u}_{px}:=(u_{1,px},u_{2,px},\ldots,u_{m,px})\\[0.3cm]
u_{j,kx}:=\frac{\p^k u_j}{\p x^k}.
\end{gather*}
Assume that (\ref{sys-F-j}) admits $m$ conserved currents, 
$\{\Phi_1^t,\Phi_2^t,\ldots,\Phi_m^t\}$,
with corresponding flux, 
$\{\Phi_1^x,\Phi_2^x,\ldots,\Phi_m^x\}$,
and the notation 
\begin{gather*}
{\bf \Phi}^t:=(\Phi_1^t,\ldots,\Phi_m^t),\qquad
{\bf \Phi}^x:=(\Phi_1^x,\ldots,\Phi_m^x).
\end{gather*}
That is
\begin{gather}
\left.\vphantom{\frac{DA}{DB}}
D_t\Phi_j^t(x,{\bf u},{\bf u}_x,\ldots)
+D_x\Phi_j^x(x,{\bf u},{\bf u}_x,\ldots)
\right|_{{\bf u}_t={\bf F}({\bf u},\,{\bf u}_x,\ldots,{\bf u}_{px})}
=0\\
j=1,2,\ldots,m.\nn
\end{gather}
We now introduce $m$ potential variables 
$\{v_1,v_2,\ldots,v_m\}$, such that
\begin{subequations}
\begin{gather}
v_{j,x}=\Phi_j^t(x,{\bf u},\,{\bf u}_x,\ldots)\\[0.3cm]
v_{j,t}=-\Phi_j^x(x,{\bf u},\,{\bf u}_x,\ldots),
\end{gather}
\end{subequations}
with corresponding potential system
\begin{gather}
\label{sys-H-j}
v_{j,t}=H_j({\bf v}_x,\,{\bf v}_{xx},\ldots,{\bf
  v}_{px})+\sum_{i=1}^m\gamma_{ij} v_i
,\qquad j=1,2,\ldots,m.
\end{gather}
Analogue to Proposition 1, we now have

\strut\hfill

\noindent
{\bf Proposition 5:} {\it
The condition on $\Phi_j^t$ which allows system (\ref{sys-F-j})
to be potentialised in system (\ref{sys-H-j}) is given by the
following conditions:
\begin{gather}
\label{sys-main-cond}
\left.
\vphantom{\frac{DA}{DB}}
D_xH_j({\bf \Phi}^t,D_x{\bf \Phi}^t,\ldots, D^{p-1}_x{\bf \Phi}^t)
+\sum_{i=1}^m\gamma_{ij}\Phi_i^t
=D_t\Phi_j^t
\right|_{{\bf u}_t={\bf F}({\bf u},\,{\bf u}_x,\ldots,{\bf
    u}_{px})}\\[0.3cm]
j=1,2,\ldots,m,\nn
\end{gather}
where $H_1,H_2,\ldots,H_m$ are given functions
and $\gamma_{ij}$ are given constants.
}

\strut\hfill

Similar to the case of scalar equations, we can define
$\bigtriangleup$-auto-B\"acklund
transformations of type I, II and III for systems of the 
form (\ref{sys-F-j}). An example of a $\bigtriangleup$-auto-B\"acklund
transformation of type I is given below.

\strut\hfill

We now consider systems of the form (\ref{sys-F-j}) that can be
potentialised or mutipotentialised in a linear system of $m$
evolution equations of order $p$,
\begin{gather}
v_{j,t}=({\cal L}_j^{(p)}[{\bm \alpha}_1^j],\,{\cal L}_j^{(p)}[{\bm
    \alpha}_2^j],\ldots,
{\cal L}_j^{(p)}[{\bm \alpha}_m^j])\,\cdot\,(v_1,v_2,\ldots,v_m)\equiv 
\sum_{k=0}^m{\cal L}_j^{(p)}[{\bm \alpha}_k^j]\,v_k\\
j=1,2,\ldots,m,\nn
\end{gather}
where ${\cal L}_j^{(p)}$ is the linear operator of order $p$
and ${\bm
  \alpha}_k^j=(\alpha_{k0}^j,\,\alpha_{k1}^j,\ldots,\alpha_{kp}^j)$
are constants. Here the linear operator ${\cal L}_j^{(p)}[{\bm
  \alpha}_k^j]$
is defined as follows:
\begin{gather}
{\cal L}_j^{(p)}[{\bm \alpha}_k^j]:=\alpha_{k0}^j\,D_x^0+\alpha_{k1}^j\,D_x^1+\cdots+\alpha_{kp}^j\,D_x^p.
\end{gather}
We consider an example of such linearisable systems.

\strut\hfill

\noindent
{\bf Example:} Let 
\begin{subequations}
\begin{gather}
\label{sys-ex1-eq-v1}
v_{1,t}=v_{1,xx}\\[0.3cm]
\label{sys-ex1-eq-v2}
v_{2,t}=v_{2,xx}
\end{gather}
\end{subequations}
and find $F_1$ and $F_2$ such that
\begin{subequations}
\begin{gather}
\label{sys-ex1-eq-u1}
u_{1,t}=F_1(u_1,u_2,u_{1,x},u_{2,x},u_{1,xx},u_{2,xx})\\[0.3cm]
\label{sys-ex1-eq-u2}
u_{2,t}=F_2(u_1,u_2,u_{1,x},u_{2,x},u_{1,xx},u_{2,xx}).
\end{gather}
\end{subequations}
The associated auxiliary system for
(\ref{sys-ex1-eq-u1})-(\ref{sys-ex1-eq-u2}) is
\begin{subequations}
\begin{gather}
v_{1,x}=\Phi_1^t(x,u_1,u_2,u_{1,x},u_{2,x},\ldots),\qquad 
v_{1,t}=-\Phi_1^x(x,u_1,u_2,u_{1,x},u_{2,x},\ldots)\\[0.3cm]
v_{2,x}=\Phi_2^t(x,u_1,u_2,u_{1,x},u_{2,x},\ldots),\qquad
v_{2,t}=-\Phi_2^t(x,u_1,u_2,u_{1,x},u_{2,x},\ldots).
\end{gather}
\end{subequations}
Following Proposition 5, condition (\ref{sys-main-cond})
reduces to
\begin{gather}
\left.\vphantom{\frac{DA}{DB}}
D_t\Phi_1^t\right|_{{\bf u}_t={\bf F}}
=D_x^2\Phi_1^t,\qquad 
\left.\vphantom{\frac{DA}{DB}}
D_t\Phi_2^t\right|_{{\bf u}_t={\bf F}}
=D_x^2\Phi_2^t
\end{gather}
We now have to make an assumption for the dependence of $\Phi_1^t$
and $\Phi_2^t$. The simplest case is
\begin{gather}
\Phi_j^t=f_j(u_1,u_2),\qquad j=1,2,
\end{gather}
where $f_1$ and $f_2$ are arbitrary functions of $u_1$ and $u_2$.
This leads to the system
\begin{gather}
\label{matrix-sys-1}
\left(
\ba{ll}
f_{1,u_1}&
f_{1,u_2}
\\
f_{2,u_1}
&
f_{2,u_2}
\ea
\right)
\left(\ba{l}
u_{1,t}\\
u_{2,t}
\ea
\right)
=\left(
\ba{l}
D_x^2f_1\\
D_x^2f_2
\ea\right).
\end{gather}
System (\ref{sys-ex1-eq-u1}) - (\ref{sys-ex1-eq-u2})
then takes the form
\begin{subequations}
\begin{gather}
u_{1,t}=
u_{1,xx}
+\left(\frac{f_{2,u_2}f_{1,u_2u_2}-f_{1,u_2}f_{2,u_2u_2}}{W}\right)u_{2,x}^2
+\left(\frac{f_{2,u_2}f_{1,u_1u_1}-f_{1,u_2}f_{2,u_1u_1}}{W}\right)u_{1,x}^2
\nn
\\[0.3cm]
\label{sys-ex1-trans-u1}
\qquad
+2\left(\frac{f_{2,u_2}f_{1,u_1u_2}-f_{1,u_2}f_{2,u_1u_2}}{W}\right)u_{1,x}u_{2,x}
\\[0.3cm]
u_{2,t}=u_{2,xx}
+\left(\frac{f_{1,u_1}f_{2,u_2u_2}-f_{2,u_1}f_{1,u_2u_2}}{W}\right)u_{2,x}^2
+\left(\frac{f_{1,u_1}f_{2,u_1u_1}-f_{2,u_1}f_{1,u_1u_1}}{W}\right)u_{1,x}^2
\nn
\\[0.3cm]
\label{sys-ex1-trans-u2}
\qquad
+2\left(\frac{f_{1,u_1}f_{2,u_1u_2}-f_{2,u_1}f_{1,u_1u_2}}{W}\right)u_{1,x}u_{2,x},
\end{gather}
\end{subequations}
where $W$ is the determinant of the matirix on the right hand side of 
(\ref{matrix-sys-1}), i.e.
\begin{gather}
\label{W-condition}
W:=f_{1,u_1}f_{2,u_2}-f_{1,u_2}f_{2,u_1}\neq 0.
\end{gather}
Hence system (\ref{sys-ex1-trans-u1}) - (\ref{sys-ex1-trans-u2})
linearises in system (\ref{sys-ex1-eq-v1}) - (\ref{sys-ex1-eq-v2})
by the relations
\begin{subequations}
\begin{gather}
v_{1,x}=f_1(u_1,u_2)\\[0.3cm]
v_{2,x}=f_2(u_1,u_2)
\end{gather}
\end{subequations}
for any differentiable functions $f_1$ and $f_2$ which satisfy
condition (\ref{W-condition}). 

In order to
construct a $\bigtriangleup$-auto-B\"acklund transformation of type I for 
system (\ref{sys-ex1-trans-u1}) - (\ref{sys-ex1-trans-u2}), we need to
find a second potentialisation for (\ref{sys-ex1-trans-u1}) -
(\ref{sys-ex1-trans-u2}) in the same system (\ref{sys-ex1-eq-v1}) -
(\ref{sys-ex1-eq-v2}). 
For this purpose we 
consider the system (\ref{sys-ex1-eq-u1}) - (\ref{sys-ex1-eq-u2})
in terms of the dependent variables $w_1$ and $w_2$, i.e.
\begin{subequations}
\begin{gather}
\label{sys-ex1-eq-w1}
w_{1,t}=G_1(w_1,w_2,w_{1,x},w_{2,x},w_{1,xx},w_{2,xx})\\[0.3cm]
\label{sys-ex1-eq-w2}
w_{2,t}=G_2(w_1,w_2,w_{1,x},w_{2,x},w_{1,xx},w_{2,xx}).
\end{gather}
\end{subequations}
and assume another
set of conserved currents for (\ref{sys-ex1-eq-w1}) -
(\ref{sys-ex1-eq-w2}), which we'll denote by $\Psi_1^t$ and
$\Psi_2^t$.
We assume the form
\begin{subequations}
\begin{gather}
\Psi_1^t=g_1(w_1,w_2)w_{1,x}+h_1(w_1,u_2)w_{2,x}\\[0.3cm]
\Psi_2^t=g_2(w_1,w_2)w_{1,x}+h_2(w_1,w_2)w_{2,x}.
\end{gather}
\end{subequations}
By Proposition 5, this leads to several systems of which we 
show here only one, namely the system
\begin{subequations}
\begin{gather}
w_{1,t}=
w_{1,xx}
+\left(\frac{h_1g_{2,w_2}-g_2h_{1,w_2}}{Q}\right)w_{2,x}^2
+\left(\frac{h_1h_{2,w_1}-g_2g_{1,w_1}}{Q}\right)w_{1,x}^2
\nn
\\[0.3cm]
\label{sys-ex1-AB-w1}
\qquad
+2\left(\frac{h_1h_{2,w_2}-g_2h_{1,w_1}}{Q}\right)
w_{1,x}w_{2,x}
\\[0.3cm]
w_{2,t}=w_{2,xx}
+\left(\frac{h_2h_{1,w_2}-g_1g_{2,w_2}}{Q}\right)w_{2,x}^2
+\left(\frac{h_2g_{1,w_1}-g_1h_{2,w_1}}{Q}\right)w_{1,x}^2
\nn
\\[0.3cm]
\label{sys-ex1-AB-w2}
\qquad
+2\left(\frac{h_2h_{1,w_1}-g_1h_{2,w_2}}{Q}\right)
w_{1,x}w_{2,x},
\end{gather}
\end{subequations}
where the following conditions must hold:
\begin{gather}
\label{conditions-w}
h_{1,w_1}-g_{1,w_2}=0,\qquad h_{2,w_2}-g_{2,w_1}=0.
\end{gather}
Here $Q$ is defined as follows:
\begin{gather}
Q:=h_1h_2-g_1g_2\neq 0.
\end{gather}
Hence system (\ref{sys-ex1-AB-w1}) - (\ref{sys-ex1-AB-w2})
linearises in system (\ref{sys-ex1-eq-v1}) - (\ref{sys-ex1-eq-v2})
by the relations
\begin{subequations}
\begin{gather}
v_{1,x}=g_1(w_1,w_2)w_{1,x}+h_1(w_1,u_2)w_{2,x} \\[0.3cm]
v_{2,x}=g_2(w_1,w_2)w_{1,x}+h_2(w_1,w_2)w_{2,x}
\end{gather}
\end{subequations}
for functions $g_1,\ g_2,\ h_1$ and $h_2$ which satisfy the conditions 
(\ref{conditions-w}). A $\bigtriangleup$-auto-B\"acklund
transformation of type I follows for system 
(\ref{sys-ex1-trans-u1}) - (\ref{sys-ex1-trans-u2})
when the systems (\ref{sys-ex1-trans-u1}) - (\ref{sys-ex1-trans-u2})
and (\ref{sys-ex1-AB-w1}) - (\ref{sys-ex1-AB-w2}) are quivalent. This
is achieved for the case
\begin{subequations} 
\begin{gather}
h_1(w_1,w_2)=\pde{f_2}{w_2},\quad
h_2(w_1,w_2)=\pde{f_1}{w_1}\\[0.3cm]
g_1(w_1,w_2)=\pde{f_2}{w_1},\quad
g_2(w_1,w_2)=\pde{f_1}{w_2}.
\end{gather}
\end{subequations}
This leads to the following

\strut\hfill

\noindent
{\bf Proposition 6:} {\it A $\bigtriangleup$-auto-B\"acklund transformation
of type I for system 
(\ref{sys-ex1-trans-u1}) - (\ref{sys-ex1-trans-u2}) is given by the
relation
\begin{subequations}
\begin{gather}
\label{sys-AB-1}
f_1(u_1,u_2)=\pde{f_2(w_1,w_2)}{w_1}w_{1,x}+\pde{f_2(w_1,w_2)}{w_2}w_{2,x}\\[0.3cm]
\label{sys-AB-2}
f_2(u_1,u_2)=\pde{f_1(w_1,w_2)}{w_1}w_{1,x}+\pde{f_1(w_1,w_2)}{w_2}w_{2,x},
\end{gather}
\end{subequations}
where $\{u_1,\ u_2\}$ and $\{w_1,\ w_2\}$ satisfy system 
(\ref{sys-ex1-trans-u1}) - (\ref{sys-ex1-trans-u2}) for any
nonconstant differentiable functions $f_1,\ f_2$ that satisfy
condition (\ref{W-condition}).}

\strut\hfill

\noindent
Fo demonstration we consider a special case of the transformation 
(\ref{sys-AB-1}) - (\ref{sys-AB-2}):
Let
\begin{gather}
f_1(u_1,u_2)=u_1u_2,\qquad f_2(u_1,u_2)=\frac{u_1}{u_2}.
\end{gather}
The relation (\ref{sys-AB-1}) - (\ref{sys-AB-2}) then reduces to
\begin{subequations}
\begin{gather}
\label{ab-u1}
u_1=\frac{1}{w_2}\left(w_2^2w_{1,x}^2-w_1^2w_{2,x}^2\right)^{1/2}\\[0.3cm]
\label{ab-u2}
u_2=\frac{1}{w_2}\frac{\left(w_2^2w_{1,x}^2-w_1^2w_{2,x}^2\right)^{1/2}}
{w_2w_{1,x}+w_1w_{2,x}},
\end{gather}
\end{subequations}
which is valid for the system
\begin{subequations}
\begin{gather}
\label{ex-ab-1}
u_{1,t}=u_{1,xx}+\left(\frac{u_1}{u_2^2}\right)u_{2,x}^2\\[0.3cm]
\label{ex-ab-2}
u_{2,t}=u_{2,xx}-\frac{1}{u_2}u_{2,x}^2+\left(\frac{2}{u_1}\right)u_{1,x}u_{2,x}.
\end{gather}
\end{subequations}
Thus for any functions, $\{w_1,\ w_2\}$, that satisfy system 
(\ref{ex-ab-1}) - (\ref{ex-ab-2}), the relation (\ref{ab-u1}) - (\ref{ab-u2})
provides a new solution $\{u_1,\ u_2\}$ for that system.

\section{The converse problem in higher dimensions}

The extension to higher dimensions is certainly a nontriavial problem. The aim
in the current paper is to propose a method of converse potentialisation for
evolution equations in $n$ dimensions in an analogue manner to
that proposed in  Proposition 1 for evolution equations in $(1+1)$
dimensions. 
We consider here the case of second-order evolution
equations and, moreoever, equations which can be potentialised in 
a linear autonomous evolution equation. 
%
In particular, we consider $n$-dimensional second-order autonomous evolution
equations in the dependent variable, $u$, and independent variables,
\begin{gather}
\{t,x,y_1,y_2,\ldots,y_{n-2}\},
\end{gather}
of the form
\begin{gather}
\label{F-n-D}
u_t=F(u,u_x,u_{xx},u_{xy_1},\ldots,u_{xy_{n-2}},u_{y_1y_1},u_{y_1y_2},\ldots,u_{y_{n-2}y_{n-2}})
\end{gather}
where $n>2$. Assume now that there exist functions,
\begin{gather}
\{\Phi^t,\Phi^x,\Phi^{y_1},\ldots,\Phi^{y_{n-2}}\}
\end{gather}
for (\ref{F-n-D}), such that
\begin{gather}
\label{Conserv-Form-n-D}
\left.
\vphantom{\frac{DA}{DB}}
D_t\Phi^t+D_x\Phi^x+D_{y_1}\Phi^{y_1}+\cdots
D_{y_{n-2}}\Phi^{y_{n-2}}\right|_{u_t=F}=0.
\end{gather}
Following \cite{Slebo} and \cite{Bluman+Kumei} we introduce $n-1$
potential variables,
\begin{gather}
\{v_1,v_2,\ldots,v_{n-1}\}
\end{gather}
and the following auxiliary system for (\ref{F-n-D}):
\begin{subequations}
\begin{gather}
\pde{v_1}{x}=\Phi^t\\[0.3cm]
\pde{v_2}{y_1}+\pde{v_1}{t}=-\Phi^x\\[0.3cm]
\pde{v_3}{y_2}+\pde{v_2}{x}=\Phi^{y_1}\\[0.3cm]
(-1)^{j-1}\left(\pde{v_j}{y_{j-1}}+\pde{v_{j-1}}{y_{j-3}}\right)=\Phi^{y_{j-2}},\quad
3<j<n\\[0.3cm]
(-1)^{n-1}\pde{v_{n-1}}{y_{n-3}}=\Phi^{y_{n-2}}
\end{gather}
\end{subequations}
We now introduce a second-order linear equation in the potential
variable $v_1$ 
and the remaining potential variables, $\{v_2,\,v_3,\ldots,v_{n-1}\}$,
in the form
\begin{subequations}
\begin{gather}
\label{n-pot-eq1}
v_{1,t}=G_{\cal L}\left(v_{1,x},\,v_{1,xx},\,v_{1,xy_1},\,\ldots,\,v_{1,xy_{n-2}}\right)\\[0.3cm]
\label{n-pot-eq2}
v_j=v_{1,x},\qquad
j=2,3,\ldots,n-1,
\end{gather}
\end{subequations}
where $G_{\cal L}$ is a linear function of its arguments, i.e.
\begin{gather*}
G_{\cal L}\equiv {\cal L}^{(2)}[{\bm \alpha},{\bm \beta}]\,v_1\\[0.3cm]
{\cal L}^{(2)}[{\bm \alpha},{\bm \beta}]:=
\alpha_1 D_x+\alpha_2 D_x^2
+\beta_1 D_x\circ D_{y_1}
+\beta_2 D_x\circ D_{y_2}
+\cdots
+\beta_{n-2} D_x\circ D_{y_{n-2}}.
\end{gather*}
Here $\alpha_j$ and $\beta_j$ are given constants. 
It is instructive to consider the cases $n=3$ separately:

\strut\hfill

\noindent
{\bf Case $n=3$:} The independent variables are $\{t,x,y_1\equiv y\}.$
The linear potential equation in $v_1$ is
\begin{subequations}
\begin{gather}
\label{n=3-pot-eq1}
v_{1,t}=G_{\cal L}(v_{1,x},\,v_{1,xx},\,v_{1,xy})\\[0.3cm]
\label{n=3-pot-eq2}
\mbox{with}\ v_2=v_{1,x}.
\end{gather}
\end{subequations}
We aim to identify the 2nd-order equation
\begin{gather}
\label{n=3-eq-u}
u_t=F(u,\,u_x,\,u_{xx},\,u_{xy},\,u_y,\,u_{yy})
\end{gather}
and $\{\Phi^t,\,\Phi^x,\,\Phi^y\}$, such that
\begin{gather}
\label{n=3-Cons}
\left.
\vphantom{\frac{DA}{DB}}
D_t\Phi^t+D_x\Phi^x+D_y\Phi^y\right|_{u_t=F}=0
\end{gather}
which potentialises in (\ref{n=3-pot-eq1}) - (\ref{n=3-pot-eq2}) with the auxiliary system
\begin{subequations}
\begin{gather}
\label{n=3-pot-aux1}
v_{1,x}=\Phi^t\\[0.3cm]
\label{n=3-pot-aux2}
v_{2,y}+v_{1,t}=-\Phi^x\\[0.3cm]
\label{n=3-pot-aux3}
v_{2,x}=\Phi^y.
\end{gather}
\end{subequations}
Applying $D_x$ on (\ref{n=3-pot-aux2}) and $D_y$ on (\ref{n=3-pot-aux3})
and using (\ref{n=3-Cons}), we obtain
\begin{gather}
\left.
\vphantom{\frac{DA}{DB}}
v_{1,xy}=D_t\Phi^t\right|_{u_t=F}
\end{gather}
which, by the use of (\ref{n=3-eq-u}) and (\ref{n=3-pot-aux1}) results
in the following condition on $\Phi^t$ and $F$:
\begin{gather}
\label{n=3-condition}
\left.
\vphantom{\frac{DA}{DB}}
G_{\cal L}\left(
D_x\Phi^t,\,D_x^2\Phi^t,\,D_x\circ D_y\Phi^t\right)=
D_t\Phi^t\right|_{u_t=F}.
\end{gather}
For a give $\Phi^t$ and $F$ which satisfy condition
(\ref{n=3-condition}), $\Phi^x$ and $\Phi^y$ can easily be 
expressed in terms of $\Phi^t$. We have

\strut\hfill

\noindent
{\bf Proposition 7:} {\it The condition on $\Phi^t$, such that
(\ref{n=3-eq-u}), viz.
\begin{gather*}
u_t=F(u,\,u_x,\,u_{xx},\,u_{xy},\,u_y,\,u_{yy})
\end{gather*}
potentialises in (\ref{n=3-pot-eq1}) - (\ref{n=3-pot-eq2}), viz.
\begin{gather*}
v_{1,t}=G_{\cal L}(v_{1,x},\,v_{1,xx},\,v_{1,xy})\\[0.3cm]
\mbox{with}\ v_2=v_{1,x},
\end{gather*}
where $G_{\cal L}$ is a linear function of its arguments,
is given by the relation (\ref{n=3-condition}), viz. 
\begin{gather*}
\left.
\vphantom{\frac{DA}{DB}}
G_{\cal L}\left(
D_x\Phi^t,\,D_x^2\Phi^t,\,D_x\circ D_y\Phi^t\right)=
D_t\Phi^t\right|_{u_t=F}.
\end{gather*}
Then
\begin{subequations}
\begin{gather}
\Phi^x=-D_y\Phi^t-G_{\cal L}\left(
\Phi^t,\,D_x\Phi^t,\,D_y\Phi^t\right)\\[0.3cm]
\Phi^y=D_x\Phi^t
\end{gather}
\end{subequations}
}

\strut\hfill

\noindent
Note that $\bigtriangleup$-auto-B\"acklund transformations can be
introduced in a similar way as for equations and systems in $(1+1)$ dimensions.

\strut\hfill

\noindent
{\bf Example:} We consider the linear potential equation
\begin{subequations}
\begin{gather}
v_{1,t}=v_{1,xx}+v_{1,x}+v_{1,xy}\\[0.3cm]
\mbox{with}\ \ v_2=v_{1,x}
\end{gather}
\end{subequations}
with the assumption
\begin{gather}
\Phi^t=f(u),
\end{gather}
where $f$ is any differetiable function of $u$. By Proposition 7 and
condition (\ref{n=3-condition}) we obtain the equation
\begin{gather}
\label{ex-n=3-u}
u_t=u_{xx}+u_{xy}+u_x+\frac{f''(u)}{f'(u)}\left(u_xu_y+u_x^2\right)
\end{gather}
and 
\begin{gather}
\Phi^x=-2f'(u)u_y-f'(u)u_x-f(u)\\[0.3cm]
\Phi^y=f'(u)u_x.
\end{gather}
As a second assumption we can consider (now in terms of the dependent
variable $q$)
\begin{gather}
\Phi^t=f'(q)q_x+\lambda f'(q)q_y+f(q)
\end{gather} 
which, upon applying Proposition 7, leads to the same equation,
(\ref{ex-n=3-u}), albeit in the variable $q$,
\begin{gather}
\label{ex-n=3-q}
q_t=q_{xx}+q_{xy}+q_x+\frac{f''(q)}{f'(q)}\left(q_xq_y+q_x^2\right),
\end{gather}
with 
\begin{subequations}
\begin{gather}
\Phi^x=-(\lambda+2)\left[
f''(q)q_xq_y+f'(q)q_{xy}+f'(q)q_y\right]
-2\lambda\left[
f''(q)q_y^2+f'(q)q_{yy}\right]\nn\\[0.3cm]
\qquad 
-f'(q)q_{xx}-f''(q)q_x^2-2f'(q)q_x-f(q)\\[0.3cm]
\Phi^y=f''(q)q_x^2+f'(q)q_{xx}+\lambda f''(q)q_xq_y+\lambda
f'(q)q_{xy}+f'(q)q_x
\end{gather}
\end{subequations}
A $\bigtriangleup$-auto-B\"acklund transformation of type I then
follows directly for (\ref{ex-n=3-u}), namely the relation
\begin{gather}
f(u)=f'(q)q_x+f'(q)q_y+f(q),
\end{gather}
where both $u$ and $q$ satisfy (\ref{ex-n=3-u}).

\strut\hfill

\noindent
Proposition 7 can readily be generalised to higher dimensions, i.e. 
the case $n\geq 4$:

\strut\hfill

\noindent
{\bf Proposition 8:} {\it The condition on $\Phi^t$, such that
(\ref{F-n-D}), viz.
\begin{gather*}
u_t=F(u,u_x,u_{xx},u_{xy_1},\ldots,u_{xy_{n-2}},u_{y_1y_1},u_{y_1y_2},\ldots,u_{y_{n-2}y_{n-2}})
\end{gather*}
with $n\geq 4$ potentialises in (\ref{n-pot-eq1}) - (\ref{n-pot-eq2}), viz.
\begin{gather*}
v_{1,t}=G_{\cal L}\left(v_{1,x},\,v_{1,xx},\,v_{1,xy_1},\,\ldots,\,v_{1,xy_{n-2}}\right)\\[0.3cm]
v_j=v_{1,x},\qquad
j=2,3,\ldots,n-1,
\end{gather*}
where $G_{\cal L}$ is a linear function of its arguments,
is given by the relation 
\begin{gather}
\left.
\vphantom{\frac{DA}{DB}}
G_{\cal L}\left(
D_x\Phi^t,\,D_x^2\Phi^t,\,D_x\circ D_{y_1}\Phi^t,\,
\ldots,\,D_x\circ D_{y_{n-2}}\Phi^t\right)=
D_t\Phi^t\right|_{u_t=F}.
\end{gather}
Then
\begin{subequations}
\begin{gather}
\Phi^x=-D_{y_1}\Phi^t-G_{\cal L}\left(
\Phi^t,\,D_x\Phi^t,\,D_{y_1}\Phi^t,\,D_{y_{n-2}}\Phi^t
\right)\\[0.3cm]
\Phi^{y_1}=D_{y_2}\Phi^t+D_x\Phi^t\\[0.3cm]
\Phi^{y_{j-2}}=
(-1)^{j-1}\left(
D_{y_{j-1}}\Phi^t+D_{y_{j-3}}\Phi^t
\right),\qquad 3< j<n,\ \ n\geq
4\\[0.3cm]
\Phi^{y_{n-2}}=(-1)^{n-1}D_{y_{n-3}}\Phi^t.
\end{gather}
\end{subequations}
}

It should be clear that Propositions 7 and 8 can be generalised to
equations of any order and converse multipotentialisations can be
introduced in
a similar way as for equations in $(1+1)$ dimensions. 
However it is also clear that the linear
equations in the form (\ref{n-pot-eq1}) - (\ref{n-pot-eq2}) is not the
most general linearisable case. 

A detailed study of the more general converse
potentialisation and converse multipotentialisation for
higher-dimensional equations 
and systems will be undertaken elsewhere.

\section{Concluding remarks}
We have introduced the converse multipotentialisation problem for equations
and systems in $(1+1)$ dimensions and also given a proposal for the
extension to higher dimensions. Triangular ($\bigtriangleup$) auto-B\"acklund
transformations were introduced and it was shown that
these transformations can be derived systematically by the converse
methodology.

The results listed in Cases I, II and III in Section 3 show that
by the converse multipotentialisation of the
linear evolution equation we were able to identify an extensive 
family of nonlinear evolution equations; all related to the 
linear evolution equation by the composition of the corresponding 
conserved currents (see Diagram 6). By systematically applying this 
converse methodology we obtained, for example, equation (\ref{main-f})
{\it viz.}
\begin{gather*}
v_t=v_{xxx}+\left(\frac{3f''}{f'}-\frac{3f'}{2f}\right)v_xv_{xx}
+\left(\frac{f'''}{f'}-\frac{3f''}{2f}+\frac{3}{4}\left(\frac{f'}{f}\right)^2\right)v_x^3
-\frac{3}{2}\alpha f v_{xx}\nn\\[0.3cm]
\qquad +\frac{3}{4}\alpha^2f^2v_x
-\frac{3}{2}\alpha 
\left(f'+\frac{ff''}{f'}\right)v_x^2.
\end{gather*}
which can be viewed as a generalised 
Calogero-Degasperis-Ibragimov-Shabat equation, (\ref{cdis}).
Note that (\ref{main-f}) admits, for arbitrarty $f(v)$, 
only one local integrating factor,
$\Lambda(x,v,v_x,\ldots)=f'(v)$
and hence only one local conservation law.
Moreover (\ref{main-f})  also includes the third-order potential
Burgers' equation, (\ref{burgers}), (for $\alpha=0$ and
$f(v)=\exp(2v)$). We consider this to be an interesting example that could 
inspire the reader to exploit this methodology to find relations between
other equations and possibly derive generalised versions of equations
that may have been introduced earlier by {\it ad hoc} methods. 

It should be clear from the examples reported here that a systematic
application of the converse methodology for equations and
systems can provide useful information regarding transformations
between equations as well as certain types of auto-B\"acklund transformations. 

The case of 
higher-dimensional equations and systems 
needs to be investigated further. We have only proposed here 
one possibility of the converse problem for higher-dimensional equations, namely the
case where the higher-dimensional equation can be linearised
in a specific type of linear equation in terms of its potential 
variables. A more detailed description of this problem is subject to 
future studies and will be presented elsewhere.


\begin{thebibliography} {99}

\bibitem{Bluman+Kumei}
Bluman G W and Kumei S
{\it Symmetries and Differential Equation}
Springer Verlag - New York, 1989

\bibitem{Calogero87}
Calogero F,
The evolution partial differential equation
$u_t=u_{xxx}+3(u_{xx}u^2+3u_x^2u) +3 u_xu^4$,
{\it J. Math. Phys.} {\bf 28} (1987), 538--555.

\bibitem{EE}
Euler N and Euler M (2001) 
A tree of linearisable second-order evolution equations by generalised
hodograph transformations, {\it J. Nonlinear Math. Phys.} {\bf 8}, 342--362. 

\bibitem{EE-RO-1}
Euler M and Euler N (2007) 
Second-order recursion operators of third-order evolution equations
with fourth-order integrating factors 
{\it J. Nonlinear Math. Phys.} {\bf 14}, 313--315. 


\bibitem{Euler_nonlocal_2009}
Euler N and Euler M (2009) 
On nonlocal symmetries, nonlocal conservation laws and nonlocal
transformations of evolution equations: Two linearisable hierarchies,
{\it J. Nonlinear Math. Phys.} {\bf 16} 489--504. 


\bibitem{Euler_NK_2009}
Euler N and Euler M (2009) 
Multipotentialisation and iterating-solution formulae: The
Krichever-Novikov  equation, 
{\it J. Nonlinear Math. Phys.} {\bf 16} Suppl., 93--106. 

\bibitem{EEP}
Euler M, Euler N, Petersson N (2003)
Linearisable hierarchies of evolution equations in (1+1) dimensions, 
{\it Stud. Appl. Math.} {\bf 111}, 315--337.

\bibitem{Ibragimov_Shabat}
Ibragimov N H and Shabat A B,
Infinite Lie-B\"acklund algebras,
{\it Funct. Anal. Appl.} {\bf 14} (1981), 313--315.

\bibitem{PEE}
Petersson N, Euler N, and Euler M (2004) 
Recursion operators for a class of integrable third-order evolution 
equations, {\it Stud. Appl. Math.} {\bf 112}, 201--225. 


\bibitem{Reyes}
Reyes E G (2005)
Nonlocal symmetries and the Kaup-Kupershmidt equation,
{\it J Math. Phys.} {\bf 46}, 073507, 19pp.

\bibitem{Slebo}
Slebodzinski W,
{\it Exterior Forms and their Applications}, 
Revised translation from the French by Abraham Goetz. Monografie
Matematyczne, Tom 52. PWN-Polish Scientific Publishers - Warsaw, 1970. 


\bibitem{SE}
W-H Steeb and N Euler,
{\it Nonlinear Evolution Equations and Painlev\'e Test},
World Scientific - Singapore, 1988.

\end {thebibliography}
\end{document}